\documentclass{aa}  
\usepackage{xcolor}

\newcommand{\radm}{{\rm rad\,m^{-2}}}
\newcommand{\dm}{{\rm pc\,cm^{-3}}}
\newcommand{\logM}{\log (M_*/{\rm M}_{\odot})}

\usepackage{siunitx}    
\usepackage{booktabs,amsmath,caption}
\usepackage{multirow}

\usepackage{natbib}
\bibpunct{(}{)}{;}{a}{}{,}

\usepackage{commath}
\usepackage{hyperref}
\usepackage{xcolor}
\hypersetup{
    colorlinks,
    linkcolor={red!50!black},
    citecolor={blue!50!black},
    urlcolor={blue!80!black}
}
\usepackage{graphicx}

\usepackage{txfonts}

 \authorrunning{Kovacs, T. O., et al.}
 \titlerunning{The halo magnetic field of a spiral galaxy at $z=0.414$}

\begin{document}

   \title{The halo magnetic field of a spiral galaxy at $z=0.414$}

   \author{Timea Orsolya Kovacs \inst{1,2,}\thanks{Corresponding author; \email{kovacs@mpia.de}}, Sui Ann Mao \inst{1}, Aritra Basu\inst{3,1}, Yik Ki Ma\inst{4,1}, B. M. Gaensler\inst{5,6,7}}

    \institute{Max-Planck-Institut für Radioastronomie, Auf dem Hügel 69, D-53121 Bonn, Germany 
    \and
    Max-Planck-Institut für Astronomie, Königstuhl 17, D-69117 Heidelberg, Germany    
    \and
    Thüringer Landessternwarte, Sternwarte 5, D-07778 Tautenburg, Germany \and
    Research School of Astronomy \& Astrophysics, Australian National University, Canberra, ACT 2611, Australia  
    \and
    Department of Astronomy and Astrophysics, University of California Santa Cruz, 1156 High St, Santa Cruz, 95064, CA, USA.
    \and 
    Dunlap Institute for Astronomy and Astrophysics, University of Toronto, 50 St. George Street, Toronto, ON M5S 3H4, Canada
    \and 
    David A. Dunlap Department of Astronomy and Astrophysics, University of Toronto, 50 St. George Street, Toronto, ON M5S 3H4, Canada}
   \date{Received 18 December 2024 / Accepted 14 July 2025}
 
  \abstract
   {}
   {Even though magnetic fields play an important role in galaxy evolution, the redshift evolution of galactic-scale magnetic fields is not well constrained observationally. In this paper we aim to provide an observational constraint on the time-scale of the mean-field dynamo, and derive the magnetic field in a distant galaxy at $z=0.414$.}
   {We obtained broadband spectro-polarimetric 1--8 GHz Very Large Array observation of the lensing system B1600+434, which is a background quasar gravitationally lensed by a foreground spiral galaxy into two images. We apply Rotation Measure (RM) synthesis and Stokes $QU$ fitting to derive the RM of the two lensed images, which we use to estimate the lensing galaxy's magnetic field.}
   {We measured the RM difference between the lensed images, and detected Faraday dispersion caused by the magneto-ionic medium of the lensing galaxy at $z=0.414$. Assuming that the RM difference is due to the large-scale regular field of the galaxy's halo, we measure a coherent magnetic field with a strength of 0.2--3.0$\,\mu$G at 0.7 kpc, and 0.01--2.8$\,\mu$G at 6.2 kpc vertical distance from the disk of the galaxy. 
   We derive an upper limit on the dynamo e-folding time: $\tau_{\rm dynamo} < 2.9~\times 10^8$~yr.  We find turbulence on scales below 50 pc, and a turbulent field strength of 0.2--12.1 $\, \mu$G.}
   {We measure the magnetic field in the halo of a spiral galaxy, and find turbulence on scales of $<50$ pc. If the RM difference is due to large-scale fields, our result follows the expectation from mean-field dynamo theory, and shows that galaxies at $z \simeq 0.4$ already have magnetic field strengths similar to present-day galaxies. However, we note the caveat of the possibility of the turbulent field of the lensing galaxy contributing to the observed RM difference.}


   \maketitle

\section{Introduction}

    Magnetic fields play an important role in the formation and evolution of galaxies \citep{2019Galax...8....4B}, as they can affect gas dynamics \citep{2023ApJ...946..114M}, galactic outflows \citep{2013ApJ...777L..38H}, propagation of cosmic rays \citep{2013ApJ...767...87W}, star formation rate (SFR, \citealt{2015MNRAS.447.3678B,2017ApJ...836..185T}) and the stellar initial mass function \citep{1987QJRAS..28..197R,2019FrASS...6....7K}. Still, the evolution of magnetic fields 
    over cosmic time is not well constrained, due to observational challenges.

    The evolution of large-scale (>1 kpc) magnetic fields is described by the mean-field dynamo theory. It can reproduce the magnetic field measurements on several kpc scales in nearby galaxies, such as magnetic pitch angle and the strength of the mean magnetic field (e.g., \citealt{2016ApJ...833...43C,2019Galax...8....4B}). Dynamo theory describes how kinetic energy is converted into magnetic energy (the amplification of the magnetic field strength), and how the turbulent magnetic field is transformed into a regular (also called coherent) magnetic field. Regular fields have a well-defined direction in the telescope beam, on kpc scales, 
    resulting in a finite average over the beam. On the other hand, turbulent fields on scales of a few 100 parsecs have spatial reversals and their average over the beam vanishes. Furthermore, turbulent fields can be either isotropic (i.e., random, have the same spatial dispersion in all directions) or anisotropic (have a preferred orientation, but reverse their direction in the telescope beam)\footnote{The different magnetic field components are described by \cite{2019Galax...8....4B}}. 
    The large-scale dynamo, which operates on kpc scale, requires the large-scale differential rotation of the galaxy's disk. Disks can already form at high redshift, such as at z$\sim$2--3 based on observations of ionized gas \citep{2006Natur.442..786G}, and at $z \simeq 4.4$ based on molecular gas \citep{2021Sci...372.1201T}. This allows the possibility of large-scale fields on scales of a few kpc existing in galaxies even at $z=2$, as there could be enough time since the formation of the disk ($z=4.4$) for the large-scale coherent field to develop \citep[$\sim$2 Gyr,][]{2009A&A...494...21A}. Modeling the dynamo consistently requires us to know the evolution of the physical parameters controlling dynamo action (e.g., angular velocity of galactic rotation, disk size, and gas density) and we need to be able to compare the derived magnetic field evolution \citep{2009A&A...494...21A,2018MNRAS.481.4410P,2022MNRAS.515.4229P} to observations of the regular magnetic field in distant galaxies.
    
    The strength and structure of magnetic fields in local galaxies can be traced by synchrotron emission (see e.g., \citealt{2019Galax...8....4B}): the total synchrotron intensity measures the plane of the sky component of the total (regular and turbulent) magnetic field, and its degree of polarization tells us about the ordered (regular and the anisotropic turbulent) magnetic field.
    However, these techniques have only been used in nearby galaxies, with current redshift limits of $z=0.009$ for individual galaxies and $z=0.018$ for interacting galaxies \citep{2013lsmf.book..215B}. The primary hurdle is the faint polarized synchrotron emission of galaxies when compared with the sensitivity of current-generation radio facilities. Even with the upcoming Square Kilometre Array (SKA) we can only realistically detect such polarized emission up to $z = 0.5$ \citep{2013pss5.book..641B}. Furthermore, the limited angular resolution causes high redshift galaxies to appear as point sources, making it impossible to study the structure of magnetic fields.
    Another way to trace the magnetic fields is by using the polarized infrared emission of dust (see e.g., \citealt{2022ApJ...936...92L} in nearby galaxies, and \citealt{2023Natur.621..483G,2024A&A...692A..34C,2025MNRAS.540L..78D} in distant galaxies). One limitation with this method is that it traces the ordered magnetic field (combination of both the regular and the anisotropic turbulent fields), thus it is unable to provide a constraint on the mean-field dynamo theory; based on dust polarization observations of the disks of nearby galaxies, it appears to originate mostly from anisotropic turbulent fields \citep{2019Galax...8....4B}. Additionally, for distant galaxies, the limited angular resolution also remains an issue, and can only be studied in the special cases of lensed galaxies \citep{2023Natur.621..483G,2024A&A...692A..34C,2025MNRAS.540L..78D}. Thus, it is difficult to obtain for a large sample.
    
    To measure the regular field, Faraday rotation experiments are needed. Only a handful of measurements exist in distant individual galaxies (\citealt{1992ApJ...387..528K,2011ApJ...733...24W,2017NatAs...1..621M}), along with some statistical detections in intervening galaxies using samples of absorption line quasar systems (\citealt{2008Natur.454..302B,2013ApJ...772L..28B,2014ApJ...795...63F,2017ApJ...841...67F,2020MNRAS.496.3142L}; and further discussed by \citealt{2018MNRAS.477.2528B}). These results suggest, galaxies up to $z=3$ may have comparable magnetic field strengths to local galaxies. \cite{2017NatAs...1..621M} demonstrated a method that can directly measure the coherent magnetic field strength in distant galaxies through observations of a polarized background quasar lensed by a foreground galaxy. By comparing the rotation measure (RM) of the lensed images of the background source, the strength and structure of the magnetic field in the lensing galaxy can be constrained. The polarized light of the background source undergoes Faraday rotation when it passes through the magneto-ionic medium along the line-of-sight, which causes the polarization angle to change as a function of wavelength, characterized by the RM. The RM ([$\radm$]) is defined as the line-of-sight ($l$ [pc]) integral of the electron density ($n_{\rm e}$ [cm$^{-3}$]) and the line-of-sight component of the magnetic field ($B_{||}$ [$\mu$G]):
    \begin{equation}
        {\rm RM} = 0.812 \int\limits_{\rm source}^{\rm observer} B_{||}(l)\, n_e(l)\, {\rm d}l.
    \end{equation}
    Because of 
    the line-of-sight integral, 
    the observed RM of the lensed images consists of the RM 
    from multiple contributors along the line-of-sight: the host galaxy of the background quasar (RM$_{\rm qso}$), the intervening galaxy of interest (RM$_{\rm gal}$), the intergalactic medium (IGM; RM$_{\rm IGM}$) and the Milky Way (RM$_{\rm MW}$). The total observed RM of the polarized light from a quasar at $z_{\rm qso}$ lensed by a galaxy at $z_{\rm gal}$ can be written as: 
    \begin{equation}
        {\rm RM}_{\rm obs} = \frac{\rm RM_{\rm qso}}{(1+z_{\rm qso})^2} + \frac{\rm RM_{\rm gal}}{(1+z_{\rm gal})^2} + {\rm RM}_{\rm IGM}(z) + \rm RM_{\rm MW},
    \end{equation}
     where RM$_{\rm qso}$ and RM$_{\rm gal}$ are rest frame RMs, and the (1+$z$)$^2$ term is the standard conversion factor to the 
     observer's frame. As lensing is achromatic \citep{1992grle.book.....S} and non-polarizing \citep{1992ApJ...390L...5D}, and assuming that the background source RM does not vary on the timescale of the gravitational time-delay, RM$_{\rm qso}$ is the same in both images. The differences in the RM contributions of the IGM \citep{2019ApJ...878...92V} and the Milky Way \citep{2012A&A...542A..93O, 2022A&A...657A..43H} are negligible between sightlines because the angular separation between the lensed images is on the scale of arcseconds or less (for estimates on these for the target in this paper, see Section \ref{Res:RMdiff_other}). As a result, any difference between the RM of the images is dominated by the magneto-ionic medium of the lensing galaxy \citep{2004JKAS...37..355N, 2017NatAs...1..621M}.

    In this paper, we estimate the magnetic field properties in the lensing system B1600+434, which has the prospect of probing the halo magnetic field strength in a galaxy at an intermediate redshift of $z=0.414$. In the top panel of Fig. \ref{fig:B1600}, we show the optical image of the system from \cite{1999AIPC..470..163K}. This system was discovered by \cite{1995MNRAS.274L..25J}, who found it to be a two-image lensing system with an image separation of $1.39\arcsec$. The redshift of the background quasar is $z$ = 1.589, and the lensing galaxy is at $z$ = 0.414 \citep{1998AJ....115..377F}. The lensing galaxy is an edge-on spiral: its axial ratio ($a/b$, where $a$ and $b$ are the major and minor axis) is 2.4 $\pm$ 0.2, and its position angle is 46$^{\circ} \pm 3^{\circ}$ \citep{1997A&A...317L..39J}, with an inclination of 81.7$^{\circ +0.4^\circ}_{\;-0.5^\circ}$ \citep{2000ApJ...533..194M}.
    The sight lines of the lensed images go through the halo, one above (image A) and the other below the disk (image B). This system complements the results of the lensing galaxy of B1152+199 from \cite{2017NatAs...1..621M}, as the two lensing galaxies are at a similar redshift, but the sightlines of B1152+199, in contrast, probe the disk of its lensing galaxy. These two systems together allow us to investigate the general magnetic field properties of a typical star-forming main sequence galaxy at $z \simeq 0.4$.
    
    This paper is organized as follows: in Section \ref{Data}, we describe the data reduction of our observations with the Karl G. Jansky Very Large Array (VLA), and in Section \ref{Results}, we show the results from RM synthesis and Stokes $QU$ fitting. In Section \ref{Discussion}, we estimate the magnetic field strength of the lensing galaxy, investigate how typical the lensing galaxy's properties are, and the implications of our findings on the large-scale magnetic field 
    amplification mechanism in galaxies. Finally, we show an overview on the typical magnetic fields of galaxies at $z \simeq 0.4$, by comparing to  B1152+199 \citep{2017NatAs...1..621M}. 
    We summarize our findings in Section \ref{Conclusion}. Throughout the paper, we use the cosmological parameters derived by \citet[][$\Omega_{\rm m}=0.3089$, $\Omega_{\rm \Lambda}=0.6911$, $H_{\rm 0} = 67.74 \,{\rm km~s^{-1}Mpc^{-1}}$]{2016A&A...594A..13P} when calculating cosmic time and length scales from redshift \citep{2006PASP..118.1711W}.

\section{Data analysis}
\label{Data}

\subsection{VLA data and imaging} 

\subsubsection{Original data}
\label{Data:or}

We performed broadband spectro-polarimetric observations over 1--8 GHz (L-, S-, and C-band) of the lensing system B1600+434 with the VLA in the A-array configuration. We chose a secondary polarization calibrator, J2202+4216, which is one of the suggested polarization calibrators for the VLA\footnote{As the primary polarization calibrators were not visible at the Local Sidereal Time (LST) range of the observation.}. However, this source can have moderate time variability and therefore does not have a standardized model of its polarization angle (PA) and polarization fraction (PF). Thus, in addition to our science observation (from here on the science dataset), we also performed an observation to obtain a model for the polarization properties of J2202+4216 (calibration dataset). The science and calibration datasets were observed on 29 October 2012 and 27 October 2012, respectively. We used the Common Astronomy Software Applications package (CASA, version 5.3.0-143, \citealt{2007ASPC..376..127M,2020ASPC..527..267E,2022PASP..134k4501C}) and used standard procedures to calibrate the data and to make image cubes of all Stokes parameters.

First, we used the calibration dataset to obtain the polarization information of J2202+4216, which we then utilized for calibrating the science dataset. In the calibration dataset, we observed 3C\,138 as the flux density and polarization calibrator, and J0319+4130 as the leakage calibrator. For the flux calibration, we used the \cite{2013ApJS..204...19P} flux density models, and the polarization model for 3C\,138 was taken from \cite{2013ApJS..206...16P}. We generated Stokes $I$, $Q$, and $U$ images of J2202+4216 with 16 MHz channel width across 1 to 8 GHz. We used the \cite{1980A&A....89..377C} deconvolver, and \cite{1995AAS...18711202B} weighting with a robust parameter of $-2$. We used the task \texttt{imfit} to fit J2202+4216 on the channel images and to determine its flux density. In Stokes $I$, we fixed the minor and major axes and the position angles for the fitted Gaussian functions to be the same as the synthesized beam, and we only let the positions and amplitudes change. The errors of the flux densities are given by the root mean square (rms) noise of a source-free region in each image. We fitted the polarization angle and the polarization fraction of J2202+4216 with a polynomial as a function of frequency, and we used this as a polarization model in the science dataset. 

For the science dataset, we used 3C\,48 as the flux density calibrator, J2202+4216 as the polarization calibrator, J2355+4950 as the leakage calibrator, and J1545+4751 as the phase calibrator. We note that we could not use 3C\,48 as a polarization calibrator for the 
entire dataset, as it has low polarization below 4\,GHz ($\lessapprox$ 3\%\footnote{\url{https://science.nrao.edu/facilities/vla/docs/manuals/obsguide/modes/pol}}), which is why we had to use J2202+4216. 
 For the flux calibration we used the \cite{2013ApJS..204...19P} models, and for the polarization calibration we used the model of J2202+4216 that we determined from our calibration dataset. As a quick independent check to see if the calibration was done correctly, we imaged 3C\,48 in Stokes $I$, $Q$, and $U$, as this source has not been used in the polarization calibration procedure. We found the PA and polarization fraction as a function of frequency to be consistent with the those from \cite{2013ApJS..206...16P}.

\begin{figure}
    \centering
    \includegraphics[width=8.6cm]{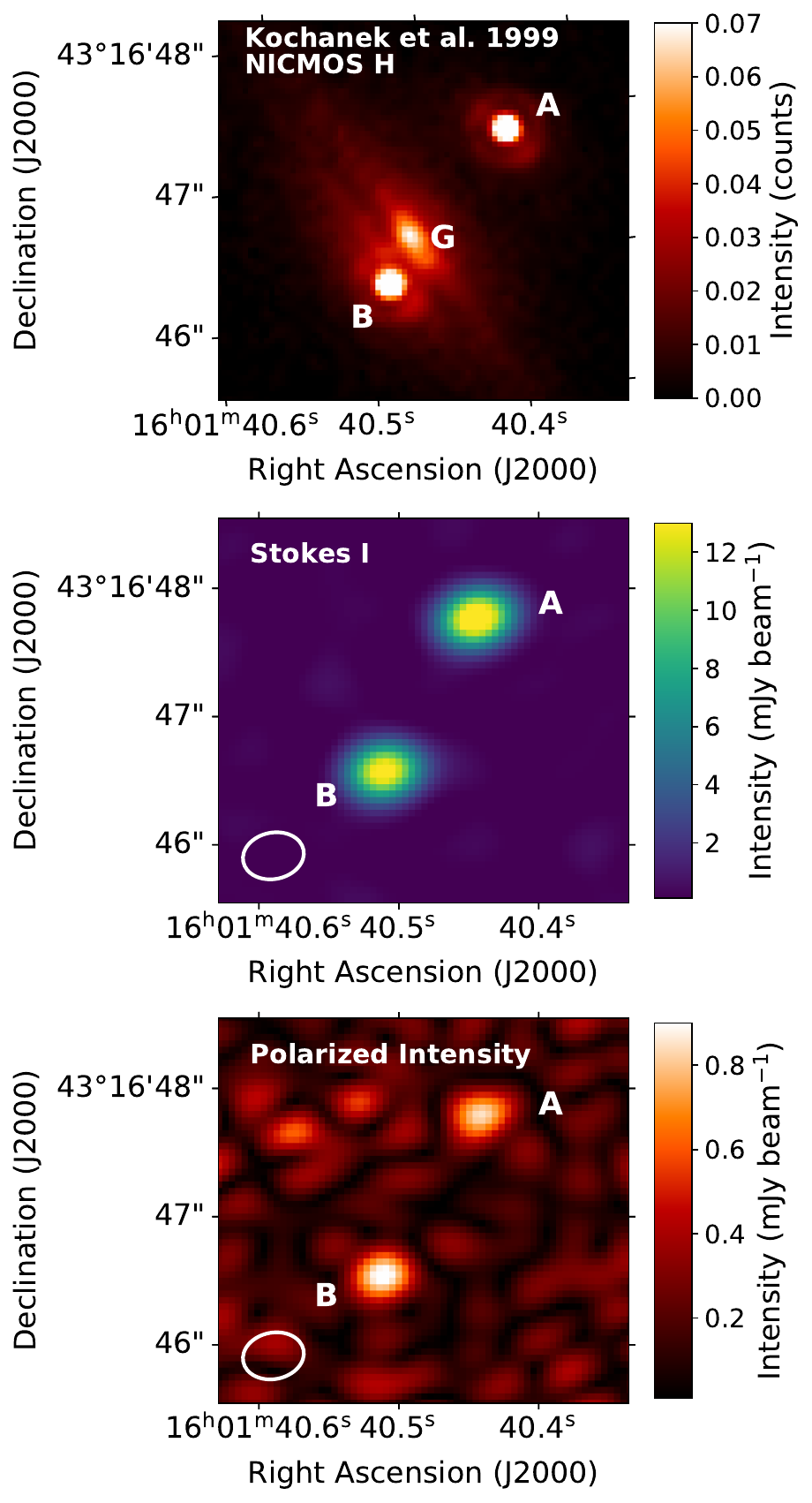}

    \caption{The Near Infrared Camera and Multi-Object Spectrometer (NICMOS) H band image (top) of the lensing system from \cite{1999AIPC..470..163K}, and the Stokes I (middle) and polarized intensity (PI, bottom) maps from our VLA data. A and B indicate the lensed images of the background quasar, and G shows the lensing galaxy. The lensing galaxy can only be seen in the NICMOS H image, and is too faint to be detected in our VLA images. The Stokes I and PI maps are 16 MHz channel maps at 3.7 GHz. The beam size is shown in the lower left corner.}
    \label{fig:B1600}
\end{figure}

We produced images with 16\,MHz channel width (see Stokes $I$ in the middle panel and the polarized intensity in the bottom panel of Fig. \ref{fig:B1600}) and derived the flux density in the same way for our target, B1600+434, in the science dataset, 
as we did for J2202+4216 in the calibration dataset. However, the lower angular resolution below 1.5\,GHz leads to the spatial blending of the two lensed images, and therefore we also fixed the positions of the fitted Gaussian functions in this frequency range, based on their positions between 1.5 and 2\,GHz. While this method works satisfactorily between 1.3 and 1.5\,GHz, we are unable to confidently separate the two lensed images and obtain accurate flux densities below 1.3\,GHz. We therefore do not include those data in our further analysis.

\subsubsection{Follow-up data in L band}
\label{Data:new}
To investigate if the variability of the background source affects the derived RM (see Section \ref{Res:bg_var})
we conducted follow-up observations of B1600+434 using the VLA in L-band (1--2 GHz) on 17 February 2021. The flagging, calibration, and imaging essentially followed the steps for the main science dataset, described in Section \ref{Data:or}, and we used 3C\,286 as the flux density and polarization calibrator, J1407+2827 as the leakage calibrator, and J1545+4751 as the phase calibrator.

   \subsection{Properties of the lensing galaxy of B1600+434}
    \label{gal_prop}
    
    To see if the lensing galaxy is a “typical” galaxy (i.e., is on the main sequence of galaxies, see e.g., \citealt{2007ApJ...660L..43N,2015A&A...575A..74S,2023MNRAS.519.1526P}), we investigated its properties.
    The lens galaxy of B1600+434 is an edge-on spiral galaxy at $z=0.414$. From lens modeling, the total mass (which includes dark matter, stars, gas, and dust) enclosed within the Einstein radius can be determined. The Einstein radius ($\theta_E$) is half of the image separation ($\theta$). In the case of B1600+434, $\theta = 1.4\arcsec$, which means the Einstein radius is 3.85 kpc at the redshift of the lensing galaxy. The enclosed total mass, log($M$/M$_{\odot}$), is estimated to be between 10.97 and 11.11 \citep{1997A&A...317L..39J,1998AJ....115..377F,1998MNRAS.295..534K}. \cite{2020A&A...634A.135G} found that according to the stellar-to-halo mass relation (derived from combining stellar mass functions from the COSMOS field with a cosmological $N$-body simulation), a halo with log($M$/M$_{\odot}$) $\sim$11 has a stellar mass of $\logM$ $\sim$8.5. As the Einstein radius does not enclose the whole galaxy, the galaxy is suspected to have a larger stellar mass than $\logM$ $\sim$8.5, suggesting it is a normal size galaxy and not a dwarf.
    
    We fitted the spectral energy distribution with CIGALE \citep{2005MNRAS.360.1413B,2009A&A...507.1793N,2019A&A...622A.103B} using 5 optical to near-infrared photometry points \citep{1997A&A...317L..39J,2000MNRAS.311..389J} to constrain the galaxy properties. We derived a stellar mass of $\logM = 10.5 \pm 0.3$ and SFR of $4.1 \pm 2.8$\,$M_{\odot}$/yr. Even though the SFR has a large uncertainty, this places the galaxy broadly at the main sequence of star forming galaxies (SFR$_{\rm MS}$ $\sim$6 $\pm$ 1 $M_{\odot}$/yr at $z\sim0.4$ for this stellar mass range, \citealt{2023MNRAS.519.1526P}), suggesting the galaxy is a normal star-forming galaxy.

\section{Results}
\label{Results}

\subsection{Rotation Measure Synthesis}

\begin{figure}
    \centering
    \includegraphics[width=8.8cm]{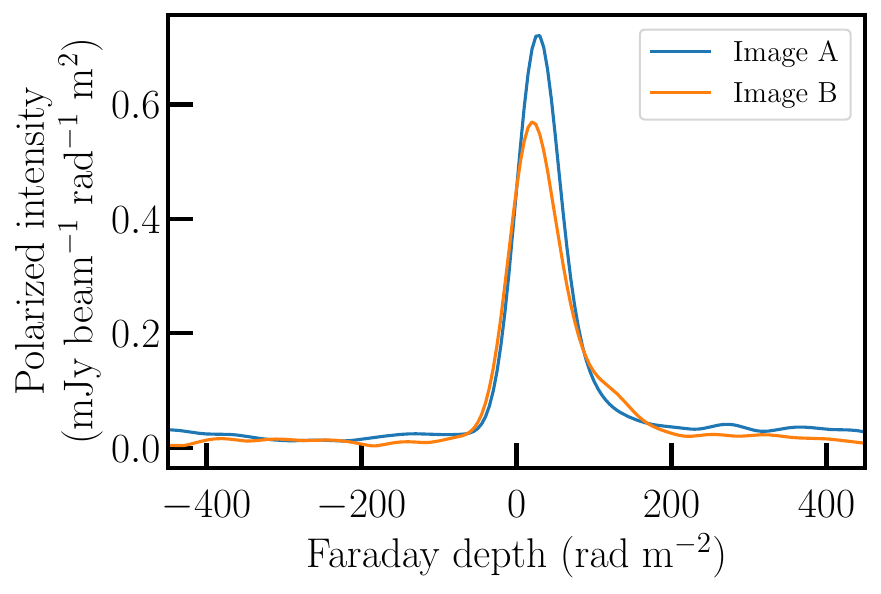}
    \caption{The clean Faraday depth spectra of image A and image B.}
    \label{fig:B1600_RMsynth}
\end{figure}

\label{results:RMsynth}

To estimate the polarization properties of the lensed images, we performed Rotation Measure synthesis \citep{2005A&A...441.1217B} on our target source, using the python package RM-Tools \citep{2020ascl.soft05003P}\footnote{\url{https://github.com/CIRADA-Tools/RM-Tools}}. RM synthesis has three main theoretical parameters that determine what we will be able to observe with our observational setup and channel width (equations 61–63 from \citealt{2005A&A...441.1217B}):
\begin{itemize}
    \item The resolution in Faraday depth space ($\delta \phi$) depends on the bandwidth ($\Delta \lambda^2 =  \lambda^2_{\rm max}- \lambda^2_{\rm min}$), where $\lambda_{\rm max}=0.231$ m and $\lambda_{\rm min}=0.037$ m, corresponding to 8 GHz and 1.3 GHz, respectively. This results in:
    \begin{equation}
    \delta \phi  = \frac{2\sqrt{3}}{\Delta \lambda^2} =  67 \, \radm.
    \end{equation}
    \item  The largest scale in Faraday depth space to which we are sensitive to with 50\%, depends on the shortest wavelength ($\lambda_{\rm min}$) of our observations:
    \begin{equation}
    \text{max-scale} = \frac{\pi}{\lambda^2_{\rm min}} = 2237 \, \radm.
    \end{equation}
    \item  The maximum observable Faraday depth ($||\phi_{\rm max}||$) depends on the channel width ($\delta \lambda$) with more than 50\% sensitivity:
    \begin{equation}
    ||\phi_{\rm max}|| = \frac{\sqrt{3}}{\delta \lambda^2} = (1\textrm{--}307) \times 10^3 \, \radm,
    \end{equation}
    where the lowest and highest values correspond to the channel widths (chosen to be 16 MHz, see Section \ref{Data:or}), which result in $\delta \lambda = 1.29 \times 10^{-3}$ m and $\delta \lambda = 5.63 \times 10^{-6}$ m at the lower and higher frequency edge, respectively.
\end{itemize}

To constrain the residual polarization leakage of our observations, we performed RM synthesis on the leakage calibrators. In the science dataset, we found the apparent polarized flux density (PI=$\sqrt{Q^2+U^2}$) of J2355+4950 to be 0.1\,mJy, and its Stokes $I$ flux density to be 1.5\,Jy. 
This corresponds to an upper limit on the residual on-axis polarization leakage of 0.01\%. The leakage calibrator J0319+4130 in the calibration dataset has an even lower residual on-axis polarization of 0.003\%. 

We perform RM synthesis on both lensed images of the target, B1600+434. In Fig. \ref{fig:B1600_RMsynth}, we show the Faraday depth spectra of the lensed images, revealing only a small difference in peak Faraday depth. We calculated the Faraday spectra between $-5000$ and $+5000\,\radm$, with a step of 5 $\radm$, and performed RM clean with a cutoff value of 0.037\,mJy (which is three times the theoretical noise level in the spectrum). We find the peaks in the Faraday depth spectra at 
${\rm RM}_{\rm A} = + 27.4 \pm 0.2 \, \radm$ and RM$_{\rm B} = + 21.0 \pm 0.4 \, \radm$ for the two lensed images.
This is calculated by RM-Tools by finding the peak and fitting a parabola to it and its two adjacent points in Faraday depth space \citep{2020ascl.soft05003P}. The Faraday depth spectra show significant deviation from a single Gaussian-like component, for example in the case of image B, there is a slight bump towards positive Faraday depth ($\sim +150\, \radm$), which could indicate a Faraday complex source.

\subsection{Stokes $QU$ fitting}
\begin{figure}
    \centering
    \begin{tabular}{c}
    \Large{Image A}\\
    \includegraphics[width=8.8cm]{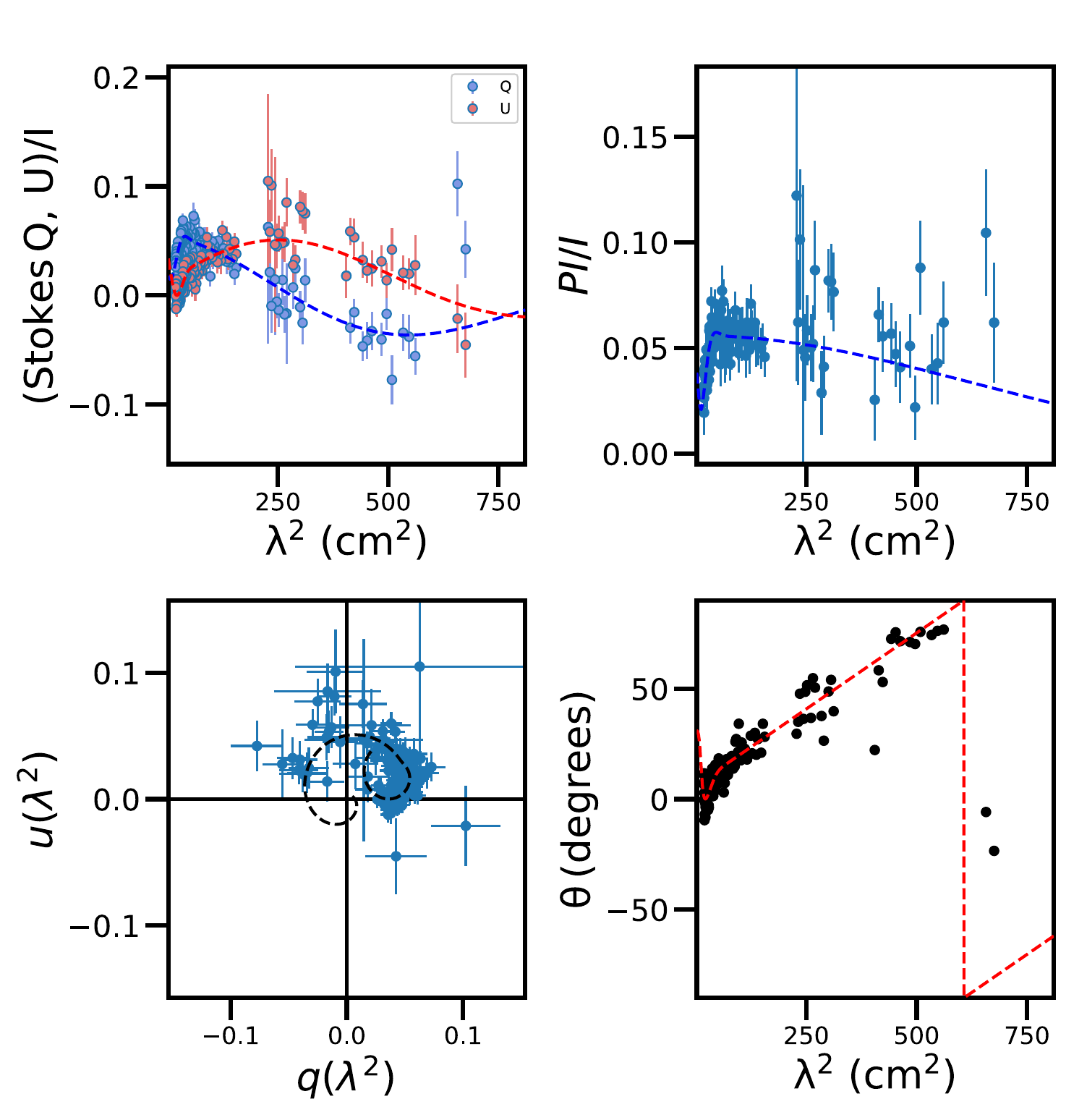}\\
    \\
    \Large{Image B}\\
    \includegraphics[width=8.8cm]{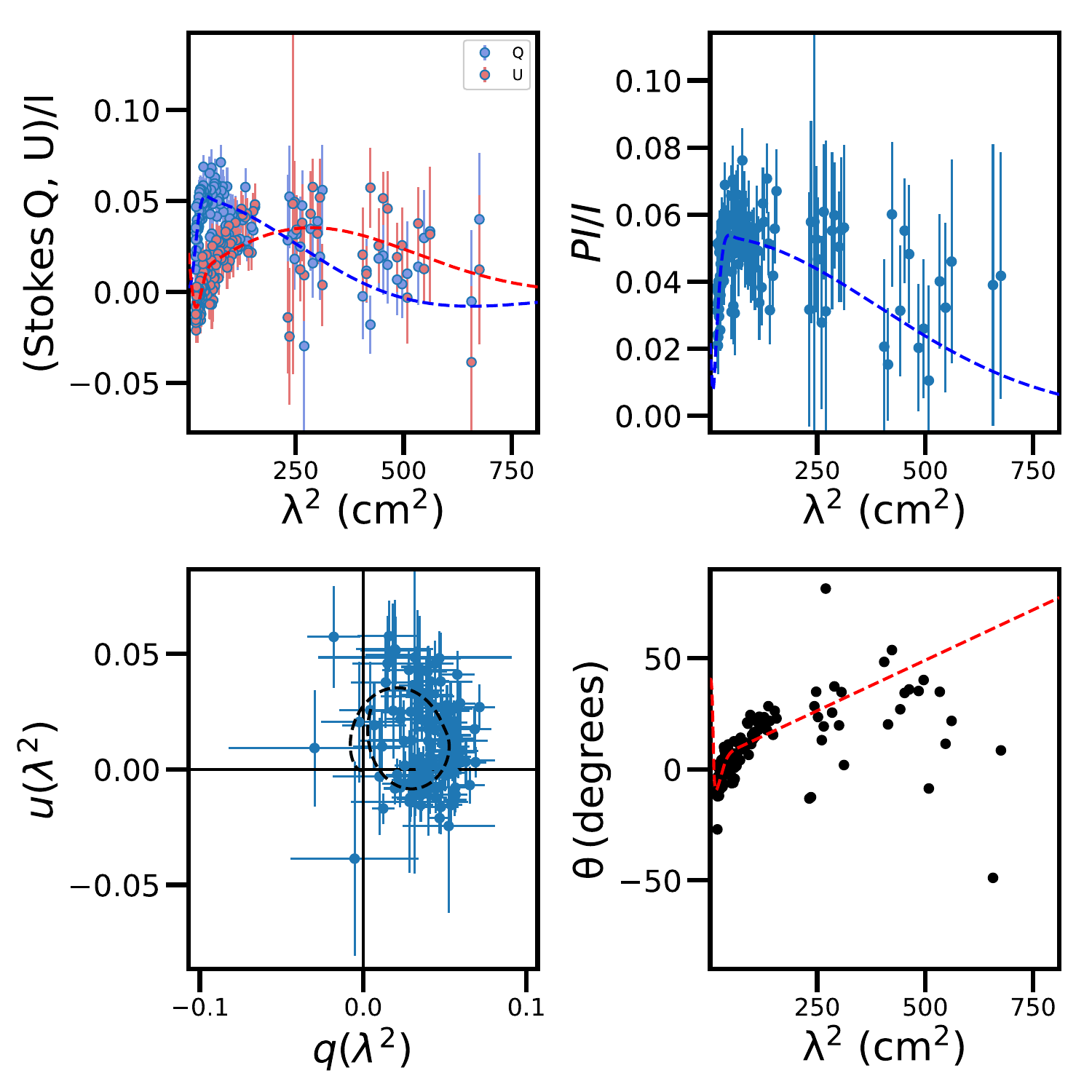}\\
    \end{tabular}
        \caption{$QU$ fitting results for image A (top panels) and B (bottom panels). Top row of each panel: the fractional Stokes parameters ($q=Q/I$, $u=U/I$), and the fractional polarized intensity (PI/$I$) as a function of $\lambda^2$. Bottom row of each panel: $u$ as a function of $q$, and the polarization angle as a function of $\lambda^2$. The fitted models are overlaid as dashed lines. }
    \label{fig:QUA}
\end{figure}

 Stokes $QU$ fitting \citep[see e.g.,][]{2011AJ....141..191F,2012MNRAS.421.3300O,2019MNRAS.487.3432M} is better at disentangling multiple polarized components than RM synthesis, as \cite{2011AJ....141..191F} previously showed.
 Thus, we use this method, and fit the fractional Stokes parameters $q = Q/I$ and $u = U/I$ as a function of frequency for both of the lensed images.

We fitted several models to the Stokes $q$ and $u$ of the lensed images independently, also considering internal and external depolarization \citep{1998MNRAS.299..189S}. One scenario is the Faraday thin model, where a synchrotron emitting source is behind a Faraday rotating screen with a homogeneous magnetic field and thermal electron density (i.e., the turbulence is on larger scales than the projected size of the background source). Another type is the \cite{1966MNRAS.133...67B} slab model with a foreground rotating screen, where the synchrotron emitting volume is also Faraday rotating.  There is also the external dispersion model, where we have a turbulent Faraday screen in the foreground (i.e., the turbulence is on smaller scales than the projected size of the background source) that causes depolarization effects. Finally, internal dispersion with foreground, where the synchrotron emitting volume is also Faraday rotating and has a turbulent magnetic field.
In the case of Faraday thin, Burn slab, external dispersion, and internal dispersion models, we tried both single and double models, where there are one and two spatially unresolved synchrotron emitting sources in the plane of the sky (i.e., polarized components), respectively. For more details on the specific models, see, for example, \cite{2019MNRAS.487.3432M}.

We chose the best model to be the one that resulted in the lowest Bayesian Information Criterion (BIC) value. A difference of more than 10 in the BIC value indicates a strong preference for a model with the lower BIC \citep{2018MNRAS.474..300S}.
The best model was one with a background source with two spatially unresolved polarized components, and two external Faraday rotating screens that have turbulence on scales smaller than the projected size of the lensed quasar. The $\sigma_{\rm RM}$ that can be measured through $QU$ fitting arises from turbulence on these small scales. However, we note that the screens can have turbulence on larger scales too, assuming a \cite{1991RSPSA.434....9K} type spectrum for the turbulent energy density with different strengths at different scales. 
This model was the best for both lensed images (with a BIC$\sim$40 lower than the next best model, double Faraday thin), shown in Fig. \ref{fig:QUA}. 
The complex polarization as a function of $\lambda^2$ is given by:
\begin{equation}
   P = {\rm PF}_{1}e^{-2\sigma^2_{\rm RM,1}\lambda^4}e^{2i({\rm PA}_{1}+{\rm RM}_1\lambda^2)}+{\rm PF}_{2}e^{-2\sigma^2_{\rm RM,2}\lambda^4}e^{2i({\rm PA}_{2}+{\rm RM}_2\lambda^2)},
\end{equation}
where $\lambda$ is the wavelength, PF$_1$ and PF$_{2}$ are the intrinsic polarization fractions, PA$_1$ and PA$_2$ are the intrinsic polarization angles, RM$_1$ and RM$_2$ are the RM values, and $\sigma_{\rm RM,1}$ and $\sigma_{\rm RM,2}$ represent the Faraday dispersions of the two components. In this model, we also assume that the polarized components have the same spectral index. The 
best-fit parameters are listed in Table~\ref{tab:StokesQU}. The presence of two polarization components is consistent with observations from the Very Large Baseline Array (VLBA) that show the substructure of the background source in Stokes\,$I$, with two sub-arcsecond scale spatial components in both the lensed images \citep{2021MNRAS.505.2610B}, that are spatially unresolved in our VLA observations.

In the rest of the paper, we refer to the polarized component with the larger PF as the main polarized component and the other as the secondary. The PA and PF values of both polarized components are consistent between images A and B, suggesting that we are 
observing the same background source and its components in both the lensed images. We find a small difference in the RM (RM$_{\rm A,1}$ vs. RM$_{\rm B,1}$, and RM$_{\rm A,2}$ vs. RM$_{\rm B,2}$) and $\sigma_{\rm RM}$ ($\sigma_{\rm RM,A,1}$ vs. $\sigma_{\rm RM,B,1}$, and $\sigma_{\rm RM,A,2}$ vs. $\sigma_{\rm RM,B,2}$) of both components between the two images. This is most likely caused by the differences in the lensing galaxy's magneto-ionic medium in the two regions probed by the lensed images. The RM of both components is larger in image A, and $\sigma_{\rm RM}$ is slightly larger in image B (although for the secondary component, the values agree within the errors). 

We try to connect the polarized components to their possible physical sources.  Active galactic nuclei (AGN) usually have larger $\vert{\rm RM}\vert$s compared to the jet \citep{2019A&A...623A.111H}. Thus, the main polarized component most likely corresponds to the jet with a lower $\vert{\rm RM}\vert$ $\sim$20$\,\radm$, and the secondary component to the emission near the core of the AGN with a high $\vert{\rm RM}\vert$ $\sim$300 $\radm$. However, it is worth noting that as the observed RM is a sum of the different components along the line of sight, a contribution of $\sim$$-150 \, \radm$ or lower would change which component has a higher $\vert{\rm RM}\vert$. We do not expect such a large contribution from the MW and the IGM (see Section \ref{Res:MW_IGM}). However, the lensing galaxy might contribute such a large RM in case of a strong magnetic field, a  larger free electron density, or a combination of both. We note that the emission from the AGN core would have a flatter spectrum than that of the jet. This would mean that the main (jet) component is mainly seen at low frequencies and the secondary (core) component at high frequencies. The large $\sigma_{\rm RM,2}$ values might be due to this effect (in contrast to being caused by a depolarizing medium), as it would result in a fit where the secondary component is suppressed at low frequency. Nevertheless, this would not affect the main component's RM difference between the lensed images, and in the rest of the paper, we do not use $\sigma_{\rm RM,2}$, so this would not affect the results of our paper.

We note that the secondary component is not seen in the clean FDF at the same RM that is derived from Stokes $QU$ fitting  (Fig. \ref{fig:B1600_RMsynth}). This is not that surprising, as, for example, \cite{2011AJ....141..191F} has shown that Stokes $QU$ fitting is better at disentangling multiple RM components than RM synthesis.

\begin{table}[]
    \caption{ Stokes $QU$ fitting results of the two polarized components of B1600+434 for the two lensed images.}
    \centering
    \begin{tabular}{ l c c}
    \hline
    \hline
         &$\text{image A}$ & $\text{image B}$ \\
         \hline
         \hline
         PA$_1$ (deg) & $6.3 \pm 1$ & \phantom{0}$3.7 \pm 1.4$\\
         PA$_2$ (deg) & \phantom{,}$71 \pm 8$ & $79 \pm 9$\\
         PF$_1$ (\%) & \phantom{0} $5.6 \pm 0.1$ &  \phantom{0}$5.4 \pm 0.2$\\
         PF$_2$ (\%) & \phantom{0} $4.2 \pm 0.6$ &  \phantom{0}$5.2 \pm 0.8$  \\
         RM$_1$ ($\radm$) & \phantom{,}$24.0 \pm 1.2$ & $15.9 \pm 2.1$ \\
         RM$_2$ ($\radm$) & \phantom{,}$342 \pm 62$ & $271 \pm 71$ \\
         $\sigma_{\rm RM,1}$ ($\radm$)& \phantom{0,}$8.4 \pm 1.5$ & $12.7 \pm 1.8$\\
         $\sigma_{\rm RM,2}$ ($\radm$)& \phantom{,}$243 \pm 26$ & $273 \pm 29$ \\
         \hline
         \hline
    \end{tabular}
    \label{tab:StokesQU}
    \tablefoot{The polarization angle (PA), polarization fraction (PF), Rotation Measure (RM) and Faraday dispersion ($\sigma$).}
\end{table}

\subsection{RM difference and possible causes}

\label{Res:RMdiff_other}

    After deriving the observed RM of the two quasar images with $QU$ fitting, we get an RM difference of $8 \pm 2\,\radm$ and $71 \pm 94\,\radm$ from the two components. 
    These results are consistent within the errors and also with 
    those obtained using RM synthesis (see Section~\ref{results:RMsynth}, Fig.~\ref{fig:B1600_RMsynth}).
    In the rest of the paper we use the result from the Stokes $QU$ fitting, and due to the large uncertainty of the RM difference from the secondary polarized component, we only consider the RM difference of the main component. 
    RM synthesis is more sensitive to the main component, and the secondary component is only captured by $QU$ fitting.
    In this section, we investigate if the RM difference can be caused by something else apart from the lensing galaxy's magneto-ionic medium. 

    \subsubsection{Contributions from the Milky Way and the Intergalactic medium}

    \label{Res:MW_IGM}
    
    According to the \cite{2022A&A...657A..43H} Galactic RM map, the Milky Way's contribution is $16.6\, \radm$ along the line-of-sight towards B1600+434, but the variation in a 10 arcmin radius around the source is only $1.4 \,\radm$, implying variation on smaller scales could be even lower. However, we note that this reflects only RM uncertainty, not the RM dispersion from the Milky Way. Extrapolated from the North Galactic Pole's RM structure-function \citep{2011ApJ...726....4S}, we expect the Galactic RM fluctuation to be $\lesssim 1.8\,\radm$ on arcsec scales. For the case of the IGM contribution, \cite{2019ApJ...878...92V} calculated the RM difference between pairs of sources that could be attributed to the IGM and found it decreases with angular separation, becoming $<1\,\radm$ on the order of arcseconds. These estimates are insufficient to explain the observed RM difference.

\subsubsection{Effects of time variability of the background source}

\label{Res:bg_var}

If there is time 
variation in the RM of the background source itself, it can lead to a contribution to the observed RM difference. This is due to the combined effect of the time variability and the time delay of the lensing system. To investigate if this can indeed be the case, we conducted follow-up observations of B1600+434 using the VLA in L-band (1--2 GHz), as described in Section \ref{Data:new}.

\cite{2021MNRAS.505.2610B} showed that B1600+434 has a time delay of $42.3 \pm 2.1$ days in radio 
(51 $\pm$ 4 days in optical, \citealt{2000ApJ...544..117B}) between the two lensed images,
and that the background source also exhibits polarization (PI, PA, and PF) variability on similar time scales. As a combined effect of time delay and time variability of the RM of the lensed quasar, the observed RM difference could also change. However, if the RM difference is mainly caused by the magneto-ionic medium of the lensing galaxy (and the properties of the galaxy's magnetic field do not vary on short time scales), the RM difference should remain constant in time. By monitoring the lensing system in polarization, we can determine the RM variability of the background source, and decide if the RM difference is caused by the galaxy.

\begin{figure*}
    \centering
    \includegraphics[width=16cm]{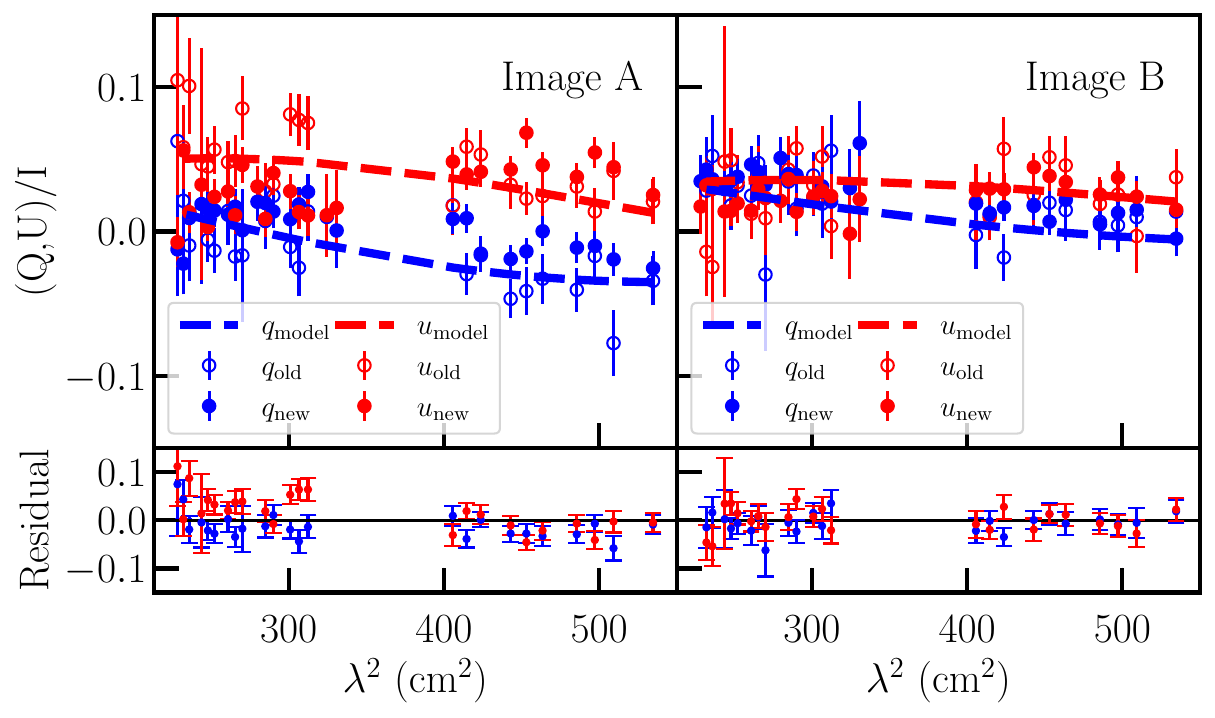}
    \caption{Comparison of our original and our follow-up data in L band. The top row shows Stokes $Q/I$ (blue) and Stokes $U/I$ (red) of our earlier data (empty circles) and our new data (filled circles) as a function of $\lambda^2$ for image A (left) and image B (right). We also plot the $QU$ fitting model (dashed lines) derived from the full frequency range of the earlier data. The bottom row shows the residual between the two datasets ($q_{\rm old}-q_{\rm new}$, $u_{\rm old}-u_{\rm new}$).}
    \label{fig:B1600_old_new}
\end{figure*}

 We applied Stokes $QU$-fitting to the new observations, and we compared the results to the $QU$ fitting of only L-band (1.3--2 GHz) of our earlier observations (see Table \ref{tab:new_QU}). In Fig. \ref{fig:B1600_old_new} we show the L-band (1.3--2 GHz) of our earlier data and our new data, over plotted on our $QU$ fitting model derived from the full frequency range (1.3--8 GHz) of the earlier data. The data points show broad agreement. We have calculated the residuals between the two datasets ($q_{\rm old}$-$q_{\rm new}$, $u_{\rm old}$-$u_{\rm new}$), separately for image A and B.
While these are close to 0, they show a slight variation between the two dates. We have also calculated the reduced $\chi^2$ of the datasets with respect to the $QU$ model: $Q$ and $U$ for image B have a $\chi^2$ lower than 2 for both the original and follow-up data. However, $Q$ and $U$ for image A in the follow-up data have $\chi^2$ of 4 and 7.5, respectively. 

We chose the model with the lowest BIC as the best-fit model for the L-band only observations (both original and follow-up): the single Faraday thin model for image A and the single external dispersion model for image B. As the L band best-fit models comprised of only one single polarized component, we concluded that we could not observe one of the polarized components found in the main science data. Based on the similarities of the polarization properties of the two lensed images in the L-band only data to the main polarized components in the full frequency range data, we argue that we found the main polarized component, but we could not observe the secondary component. This is supported by how the secondary component most likely could only be observed with higher frequency data due to its large $\sigma_{\rm RM}$ (or due to its flat spectrum). Even though the observations were taken nine years apart, the polarization properties of the lensed images do not show significant variation. We conclude that the RM of the main polarized component has not changed significantly over time (i.e., it is consistent within error), and therefore, likewise for the RM difference. This suggests that the RM difference between the images reflects the magneto-ionic medium of the foreground galaxy (i.e., there is no significant contribution by the RM time variability of the background source). However, note that our follow-up observations have higher uncertainties since we only used a narrower frequency range, which reduces the precision in Faraday space.

\begin{table}[]
    \centering
    \caption{The $QU$ fitting result of the two images of B1600+434 in L band (original and new data).}
    \begin{tabular}{l c c}
    \hline
    \hline
         & $\text{image A}$ & $\text{image B}$ \\
         \hline
         \hline
         PA$_{\rm new}$ (deg) & \phantom{,}$-0.7 \pm 6.9$ & $-10.7 \pm 6.8$\\
         PF$_{\rm new}$ (\%)  & \phantom{0,} $3.6 \pm 0.2$ & \phantom{00} $4.8 \pm 0.6$\\
         RM$_{\rm new}$ ($\radm$)& \phantom{,} $19.6 \pm 3.2$ & \phantom{0,}$17.1 \pm 3.3$\\
         $\sigma_{\rm RM, new} (\radm)$  & \phantom{0,}0 & \phantom{00} $9.2 \pm 2.5$\\
        \hline
         PA (deg) & $14.5 \pm 8.2$ & \phantom{,} $-0.4 \pm 12.7$\\
         PF (\%) & \phantom{,} $5.1 \pm 0.4$  & \phantom{0}$5.1 \pm 1.2$ \\
         RM ($\radm$)& $19.6 \pm 3.7$ & $14.6 \pm 6.4$ \\
         $\sigma_{\rm RM} (\radm)$ & \phantom{,} 0 & $10.5 \pm 4.0$ \\
         \hline
         \hline
    \end{tabular}
    \tablefoot{The new observations in L band are indicated with new in subscript.}
    \label{tab:new_QU}
\end{table}

    \subsubsection{Gravitational lensing}
    Gravitational lensing only introduces additional changes in the polarization angle in the case of a rotating or non-spherical lens. 
    This effect is less than 0.1 degrees if the image is $r=100~m$ away from the lens \citep{1992ApJ...390L...5D}, where $m$ is the geometrized mass (in geometrized unit system). In the case of a galaxy with a stellar mass of $\logM=12$, $m$ is 0.05 pc, and $r=5$ pc, which is significantly less than the distance of the lensed images from the lensing galaxy (0.7 and 6.2 kpc). This means we do not expect a change in the polarization angle between the lensed images of the background quasar due to gravitational lensing.

A change in the polarization angle can also be due to birefringence caused by axion-like particles \citep{2021PhRvL.126s1102B}. However, this effect is achromatic, thus does not cause a difference in RM.

    \section{Discussion}
    \label{Discussion}

\subsection{The rest frame RM difference}
    After excluding the possibility of other causes in Section \ref{Res:RMdiff_other}, we proceed with attributing the RM difference between the lensed images to the magneto-ionic medium of the lensing galaxy.
    To correct for the redshift dilution, we convert the observed RM (RM$_{\rm obs}$) of the two quasar images (RM$_{\rm A, obs}$= 24 $\pm$ 1 $\rm rad$ $\rm m^{-2}$, RM$_{\rm B, obs}$ = 16 $\pm$ 2 $\rm rad$ $\rm m^{-2}$) to rest frame RM ($\Delta$ RM$_{\rm rf}$):
    
    \begin{equation}
    \Delta{\rm RM}_{\rm rf} = (1+z_{\rm gal})^2 \cdot {({\rm RM_{\rm B, obs}} - {\rm RM_{\rm A, obs}})},
    \end{equation}
    where $z_{\rm gal}=0.414$ is the redshift of the lensing galaxy.
    The rest-frame RM difference is related to the electron column density ($N_{e} = \int n_e {\rm d}l $ ) and the line of sight magnetic field component ($B_{\rm ||}$) in the following way:
    \begin{equation} \label{eq:DRM}
        \Delta {\rm RM_{\rm rf}} =  0.812\,(N_{\rm e,B} B_{\rm ||,B}  - N_{\rm e,A}  B_{\rm ||,A}),
    \end{equation}
    where the quantities with subscripts A and B correspond to the positions probed by sightlines A and B, respectively. Eq. \ref{eq:DRM} assumes that B$_{||}$ is constant along the LOS, the path lengths are the same in both sightlines, and 
    the magnetic field and the electron density are uncorrelated. We derive $\Delta {\rm RM_{\rm rf}}=16 \pm 4 \,\radm$.

    In the following, we investigate if the small-scale or large-scale field of the lensing galaxy is responsible for the RM difference. For both cases, we need to first make assumptions on the electron density of the galaxy. The ranges of parameters we use in our calculations are listed in Table \ref{tab:parameter_ranges}.

    \subsection{Assumptions on electron density}
    \label{Dis:ne}
    \subsubsection{General assumptions}
    \label{Dis:ne_gen}
    This lensing system has been observed with the {\it Chandra X-ray Observatory}, from which \cite{2005ApJ...625..633D} derived the differential hydrogen column density based on X-ray absorption ($\Delta\,N_{\rm H} = N_{\rm H,B} - N_{\rm H,A} \simeq 3 \cdot 10^{21} \rm cm^{-2}$). This can be converted into differential electron column density ($\Delta\,N_{\rm e}$) by assuming an ionization fraction (f$_{\rm ion}$) lying in a wide range over $5 \textrm{--} 50\%$, we get $\Delta\,N_{\rm e}=N_{\rm e,B}-N_{\rm e,A}=48.6\textrm{--}486.1\,\dm$. The advantage of this  method is that we do not need to assume a path length through the lensing galaxy. This can be directly used in a constant (with distance from the galaxy's plane) and antisymmetric magnetic field geometry model (see Section~\ref{section:Dis_B}, item 2) to get a $B$ field strength estimate, as we only need the $\Delta N_{\rm e}$. However, this assumption is insufficient for other magnetic field geometry models (see Section \ref{section:Dis_B}, items 3--6), as instead of $\Delta N_{\rm e}$ we need the sum of the electron column densities. Thus, for these models we need to make another assumption about the electron column density of sightline A. We assume a wide range of $1 \textrm{--} 100 \,\dm$ for $N_{\rm e,A}$, which corresponds to $n_{\rm e} = 10^{-4} \textrm{--} 10^{-2}\, {\rm cm}^{-3}$ on average in a 10 kpc sightline. This ranges from approximately 0.5 to 50 times of the MW value at a vertical distance of $h_{\rm A}=6.2$\,kpc from the mid-plane of the galaxy (see Section \ref{Dis:ne_MW}). This results in $N_{\rm e,B}$~=~$N_{\rm e,A}$~+~$\Delta\,N_{\rm e}$~=~49.6--586.1 $\dm$, and $N_{\rm e,A} + N_{\rm e,B} = 50.6 \textrm{--} 686.1 \, \dm$. For all assumed values above, we consider the wide range for parameters $f_{\rm ion}$ and $N_{\rm e,A}$, to take into account the possibility that high redshift galaxies have different properties than local ones.

    \subsubsection{Electron density in MW-like galaxies}
    \label{Dis:ne_MW}
    We also investigate the scenario where we assume Milky Way like properties for the lensing galaxy (based on our findings in Section \ref{gal_prop}).

    We estimate the electron column densities by using the electron density profile derived by \cite{2020ApJ...897..124O} for the Milky Way halo:
    \begin{equation}
     n_e = n_0 \exp(-h/h_{\rm scale}), 
    \end{equation}
    where $n_0$ = 0.015 $\pm$  0.001 cm$^{-3}$, $h_{\rm scale}$ = 1.57$^{+0.15}_{-0.14}$ kpc and $h$ is the vertical distance from the Galactic mid-plane. If we assume the galaxy is completely edge-on, and we calculate with an integral path length of 10 kpc through the halo (e.g., see \citealt{2022A&A...658A...7B}), we derive electron column densities of $N_{\rm e,A} = 2\textrm{--}4 \, \dm$ and $N_{\rm e,B} = 92 \textrm{--} 100 \, \dm$ at a distance from the plane of the galaxy at $h_{\rm A}$ = 6.2 kpc (image A) and $h_{\rm B}$ = $|-0.7|$ kpc (image B), respectively. This gives us a differential electron column density $\Delta\,N_{\rm e} = 88 \textrm{--} 98 \, \dm$, which is consistent with using $f_{\rm ion} = 9 \textrm{--} 10\%$ to calculate $\Delta\,N_{\rm e}$ from $\Delta\,N_{\rm H}$. Which in turn is in agreement with the ionization fraction of the Galactic interstellar medium (10$^{+4}_{-3}$\%, \citealt{2013ApJ...768...64H}, derived from the dispersion measure and X-ray column density of radio pulsars). The sum of the electron column densities is $N_{\rm e,A} + N_{\rm e,B} = 94\textrm{--}104 \, \dm$.

    Our calculations based on \cite{2020ApJ...897..124O} suggest that $N_{\rm e,A}$ is more likely to be much less than 100 $\dm$, unless $n_e$ is an order of magnitude higher in galaxy halos at $z\sim0.4$ compared to the Milky Way, in contrast to our general assumptions. Based on spectral line observations of high redshift galaxies \citep{2021ApJ...909...78D,2023ApJ...956..139I}, $n_e$ could be twice as large in galaxies at $z\simeq0.4$ compared to $z=0$. This would result in $N_{\rm e,A} = 4\textrm{--}8 \, \dm$ and $N_{\rm e,B} = 183 \textrm{--} 200 \, \dm$, and corresponds to an ionization fraction of 18--20 \%.
    
    The ranges of electron density from the general assumptions in Section \ref{Dis:ne_gen} overlap with these but have a wider range; thus we decided to calculate the magnetic field properties with the widest (full parameter range) and narrowest (MW-like parameters) range of parameters.

    \subsection{Turbulent magnetic field}

    We detect differential Faraday dispersion, which is due to the amount of turbulence in the lensing galaxy being different in the two sightlines on scales smaller than the projected size of the background source. The differential Faraday dispersion is defined as:
    \begin{equation}
    \Delta\sigma_{\rm RM, rf}~=~(1+z_{\rm gal})^2\sqrt{\sigma_{\rm RM,1,B}^2 -\sigma_{\rm RM,1,A}^2}~=~19 \pm 5~\radm.
    \end{equation}
    From $\Delta\sigma_{\rm RM, rf}$,  we can derive an estimate on the small-scale random field (following \citealt{2017NatAs...1..621M}, and based on \citealt{2015A&ARv..24....4B}), on scales below $\sim$50 pc:\footnote{The projected size of the background source at the redshift of the lens, based on VLBI imaging \citep{2021MNRAS.505.2610B}}
    \begin{equation}
        B_r = \frac{\sqrt{3fN\Delta \sigma_{\rm RM}^2}}{0.812 \sqrt{N_{\rm e,B}^2 -N^2_{\rm e,A}}},
    \end{equation}
    where $f=0.02 \textrm{--} 0.14$ is the filling factor and $N=10\,000$ is the number of turbulent cells in the sightline (in the case of $l_1 = 10$ pc cells through a $L = 10$ kpc sightline). Using MW-like $N_e$ estimates (see Section \ref{Dis:ne}), we derive $B_r$ $\sim$1.4--6.9 $\mu$G, while considering the full parameter range $B_r$ can vary between 0.2 and 12.1 $\mu$G. We note that apart from $f_{\rm ion}$ and $N_{\rm e,A}$, the derived strengths also depend strongly on the assumptions of $N$ and $f$. 
    
    Assuming that the magnetic energy power spectrum is described by a power-law, with a slope similar to Kolmogorov turbulence, we can estimate magnetic field strengths on larger scales by extrapolating the strength measured on $<50$ pc scales. Thus, we investigate if turbulent fields can cause the RM difference we see between the two lensed images. For this, we extrapolate the turbulent magnetic field and $\sigma_{\rm RM}$ to scales of $l_2 = 1000$ pc (outer scale of turbulence in the halo of M51, \citealt{2015ApJ...800...92M}), detailed in Appendix \ref{app:turbulent}. We assume Kolmogorov spectrum for the turbulence, and we include an additional factor because we are dealing with turbulence in the halo and not in the disk. We derive $\sigma_{\rm RM}$ = 41 $\pm$ 11 $\radm$ on scales of 1000 pc, in the rest frame. The parameters that can affect our derived value significantly (i.e., by a factor of 2 if changed by a factor of 2) are the assumed ratio of the halo and disk density, and the cell size of the turbulence $l_{1}$ and $l_{2}$.

    The derived $\sigma_{\rm RM}$ on kpc-scale is of a similar magnitude as the RM difference between the images, which indicate turbulent field being a possible contributor to the measured RM difference. We stress again that extrapolating $\sigma_{\rm RM}$ to kpc scales heavily depends on the assumed ISM properties. Thus, we note that the possibility of the turbulent field contributing to the RM difference is a caveat of our method, which can only be completely excluded by joint analysis of more lensing systems (see Section \ref{Conclusion}).

    \subsection{Large-scale magnetic field geometries and strengths}
    \label{section:Dis_B}
    \subsubsection{General assumptions}
    \begin{table*}[]
        \centering
         \caption{The derived magnetic field strengths for different magnetic field geometry models. }
       \resizebox{0.85\textwidth}{!}{
        \begin{tabular}{c c c|c c c | c c c }
        \hline \hline 
             Model & Scaling & Symmetry & \multicolumn{3}{c|}{Full parameter range} &\multicolumn{3}{c}{MW-like parameters}\\
             & & & B$_0$  & B$_{\rm A}$ & B$_{\rm B}$ & B$_0$  & B$_{\rm A}$ & B$_{\rm B}$ \\
             & & & ($\mu$G) & ($\mu$G)& ($\mu$G) & ($\mu$G) & ($\mu$G)&  ($\mu$G)\\             
        \hline \hline 
             Toroidal & Constant & & [0.04, 0.4] & [0.04, 0.4] & [0.04, 0.4] & [0.15, 0.3] & [0.15, 0.3] & [0.15, 0.3] \\
             \hline
             \multirow{7}{*}{Poloidal} & Constant & anti & [0.3, 3.0] & [$-$3.0, $-$0.3] & [$-$3.0, $-$0.3] & [1.3, 1.8]  & [$-$1.8, $-$1.3]  & [$-$1.8, $-$1.3]\\  
             
              & Constant & sym. & [0.2, 2.8] & [0.2, 2.8] & [$-$0.2, $-$2.8] & [1.2, 1.7] & [1.2, 1.7] & [$-$1.2, 1.7]\\
              & $B/h$ & anti & [0.2, 2.0] & [0.02, 0.3] & [0.2, 2.9] & [0.9, 1.2] & [0.1, 0.2] & [1.2, 1.7]\\
              & $B/h$ & sym. & [0.2, 2.0] & [0.03, 0.3] & [$-$2.9, $-$0.2] &[0.9, 1.2] & [0.1, 0.2] & [$-$1.7, $-$1.2]\\
              &Exponential & anti & [0.2, 4.1] & [0.01, 2.2] & [0.2, 2.9] & [1.4, 2.5] & [0.1, 0.9] & [1.2, 1.7]\\
              & Exponential & sym. & [0.3, 4.1] & [0.01, 2.2] & [$-$2.9, $-$0.2]  & [1.3, 2.5] & [0.1, 0.9] & [$-$1.8, $-$1.2]\\
             \cline{2-9}
              & All (|$B$|) & &[0.2, 4.1] & [0.01, 2.8] & [0.2, 3.0] & [0.9, 2.5] & [0.1, 1.8] & [1.2, 1.8]\\
             \hline
        \end{tabular}}
        \tablefoot{Image A is 6.2 kpc and image B is $-0.7$ kpc away from the plane of the galaxy.}
        \label{tab:B_field_values}
    \end{table*}

    In this section, we assume that the regular field is responsible for the RM difference between the two lensed images. We derive the magnetic field strength of the lensing galaxy using simplified magnetic field geometry models (listed in Appendix \ref{app:geometry}), ranging from simple constant fields to fields that scale with distance from the galaxy's plane.

    An additional important parameter needed to derive the large-scale regular field strength is the galaxy's inclination, and (in some of our models) the distance of the lensed images from the midplane.   \cite{2000ApJ...533..194M} derived an inclination ($i$) of 81.7$^{\circ +0.4^\circ}_{\;-0.5^\circ}$ 
    by fitting elliptical isophotes to a \textit{Hubble} Space Telescope (HST) image of the galaxy. We consider image A to be above the disk with a vertical distance from the midplane ($h=0$ kpc) of $h_{\rm A} = + 6.2$ kpc and image B to be below the disk with a vertical distance from the midplane of $h_{\rm B} = - 0.7$ kpc.

    The halo field is most likely dominated by poloidal fields, which are best approximated by vertical magnetic fields (i.e., fields perpendicular to the galaxy's disk) in the inner halo (i.e., at small radii). In this case, the line-of-sight magnetic field component is given by 
    \begin{equation}
        B_{\rm || } = B_{\rm vert} \cos (i).
    \end{equation} 

    In summary, we find that if we use models that only assume a vertical field, the magnetic field strength $B_0$ is between 0.2 and 4.1 $\mu$G. We show the derived magnetic field strengths both at position A and B in Table \ref{tab:B_field_values}.

    However, it is possible that instead of a poloidal field, there is a dominating toroidal field in the halo. If there is a large-scale dynamo in the galaxy, a toroidal magnetic field will be produced in the disk, which can also be present in the halo due to advection from galactic winds \citep[see e.g.,][]{2019Galax...7...54K,2021A&A...649A..94M}. In this case, the RM we see could be mainly due to the toroidal field. Then, the line-of-sight component is given by 
    \begin{equation}
    B_{||} = B_{t} \sin(i),        
    \end{equation} where B$_t$ is the strength of the toroidal magnetic field, which we assume to be constant. We derive a field strength between 0.04 to 0.4 $\mu$G, lower than the strength resulting from assuming a poloidal field in most cases.

    We obtained such a wide range of magnetic field strength values because of the wide range of values for the parameters $f_{\rm ion}$ and $N_{\rm e,A}$. In Appendix \ref{app:parameters} and \ref{app:Bmodel}, we explore in detail how each parameter, and the choice of the magnetic field geometry model, affect the results. In summary, the assumed ionization fraction and the inclination of the galaxy  affect the derived magnetic field strengths the most, by a factor of $\sim$10 and $\sim$5, respectively. In the case of the position probed by image A, at 6.2 kpc from the midplane of the galaxy, the effect of the scale height is also significant (a factor of $\sim$10)

    We find that all models give similar ranges of magnetic field strength, apart from the toroidal model, which results in low field strengths. 
    However, we can not rule out any of the models above based on our data, as all of them give reasonable results compared to nearby galaxies and the Milky Way (see Section \ref{dis:nearby}). 

    We note that the sightlines of the lensed images might go through different heights from the plane of the galaxy, compared to the apparent distance,  due to the galaxy's finite inclination. This causes them to go through regions with different magnetic field strengths, proportional to the height from the galaxy's plane. To estimate the effect of this, we calculated the possible heights in a 10 kpc long sightline ($h_{\rm A} = $ 5.5 to 7 kpc and $h_{\rm B} = $ 0.02 to 1.4), and found the inferred average magnetic field strengths would change by 30--40\%. 
    
    \subsubsection{Large-scale magnetic field strength in the case of MW-like properties}
    
    Here we investigate 
    our results for a restricted range of parameters corresponding
    to MW-like galaxies and electron column densities ($\Delta N_{e} = 68\textrm{--}136 \,\dm$) calculated in Section \ref{Dis:ne} based on the $n_e$ profile from \cite{2020ApJ...897..124O}, and assuming a scale height of $H_{\rm halo}=2\textrm{--}8$ kpc for the exponential magnetic field geometry \citep{2019Galax...8....4B}.
    We investigated the different geometry models for this parameter case and list the results in Table \ref{tab:B_field_values}. The exponential model is the most realistic model based on what we see in nearby galaxies, for example the synchrotron emission shows exponential dependence with distance from the plane, which is also indicative of the magnetic field \citep[see e.g., ][]{2015A&ARv..24....4B, 2018A&A...611A..72K}. This results in an absolute vertical magnetic field strength of 1.2--1.8 $\mu$G at 0.7 kpc away from the plane of the disk ($B_{\rm B}$) and 0.1--0.9 $\mu$G at 6.2 kpc ($B_{\rm A}$).

    \subsubsection{Implications on the field amplification mechanism in the lensing galaxy of B1600+434}
    The mean-field galactic dynamo has a theoretical amplification timescale (e-folding time) of $\sim$2.5$\times 10^8$ yr \citep{2009A&A...494...21A}, for a disk galaxy with an average angular velocity of $\Omega = 20$ km s$^{-1}$ kpc$^{-1}$. This timescale describes the time needed to amplify a regular field that is spatially ordered on the scale of a few kpc. However, the time needed for a large-scale field to be ordered on the scale of the entire galaxy disk (ordering timescale, 10s of kpc) without reversals is only a few times shorter than the lifetime of a galaxy (i.e., on order of magnitude larger, \citealt{2009A&A...494...21A}).
    
    Present day magnetic field strengths of the order of $\mu$G are expected from the mean-field dynamo theory \citep{1988ASSL..133.....R}. We find that the large-scale magnetic field has already built up to present-day magnitude at $z=0.414$ (see Section~\ref{dis:nearby}). The measured magnetic field strength can constrain the amplification timescale of the large-scale coherent magnetic field of galaxies. To derive the e-folding time, we use the following equation (similar to \citealt{2017NatAs...1..621M}): 
    \begin{equation}
        \tau_{\rm dynamo} [{\rm Gyr}] = \frac{\Delta t}{\ln(B_{\rm disk, z=0.414}/B_{\rm seed})},
    \end{equation}
    where $B_{\rm seed}$ is the seed magnetic field of the mean-field dynamo, $\Delta t=5.97$~Gyr is the time that passed between when the disk settled into equilibrium ($z=2$, e.g., \citealt{2006Natur.442..786G}) and the redshift of the lensing galaxy ($z=0.414$). Furthermore, we assume that $B_{\rm halo}$ = 0.4 $B_{\rm disk}$ \citep{2019Galax...8....4B}\footnote{From assuming the density and energy density ratio of the disk and the halo of galaxies.}, where $B_{\rm disk, z=0.414}$ is the disk field strength at $z=0.414$. While this relation was derived for the equipartition magnetic field strength, we argue that it is also true for the regular field, based on observations of M\,51 \citep[][who derived a ratio of 0.3]{2014A&A...568A..83S}. We take the lowest magnetic field strength value we derived in the MW-like parameter case, so $B_{\rm halo} = $ 0.1 $\mu$G. We also assume that the magnetic field has already saturated, at the latest, at $z = 0.414$ (as the derived strengths are consistent with nearby galaxies, see Section \ref{dis:nearby}). 

    There are multiple seed field origin scenarios, mainly divided into primordial fields generated in the early Universe and fields generated during galaxy formation \citep{2019Galax...7...47S,2019Galax...8....4B}. The former can be estimated by measurements of cosmic voids and cosmic web filaments. Following the assumption in \citep{2017NatAs...1..621M}, $B_{\rm seed}$ would be $\geq 3 \times 10^{-16}$~G \citep[][based on cosmic voids]{2010Sci...328...73N}. Using this value, we derive an upper limit on the dynamo e-folding time: $\tau_{\rm dynamo} $\textless2.9~$\times$~10$^8$~yr, which is consistent with the theoretical value \citep{2009A&A...494...21A}. If $B_{\rm seed}$ is higher $\tau_{\rm dynamo}$ will be higher too, such as $\sim 7-8 \times 10^8$~yr for $0.04-0.11~ {\rm nG}$, \citep[][from cosmic filaments]{2023MNRAS.518.2273C}. However, if $B_{\rm seed}$ is significantly lower (e.g., $10^{-21}$ from Biermann battery origin, \citealt{2019Galax...7...47S}), it would result in $\tau_{\rm dynamo} < 1.8 \times 10^8$~yr, which is lower than the theoretical amplification timescale \citep{2009A&A...494...21A}. In the case of a seed field from the small-scale dynamo, the field strength could be of $\mu$G strength, which would result in a much larger possible e-folding time. For example, if we assume a $\sim$0.1 $\mu$G seed field, we would derive $\tau_{\rm dynamo}< 6.5 \times 10^9$~yr. We note that the apparent relation of a smaller $B_{\rm seed}$ leading to a better constraint of $\tau_{\rm dynamo}$ is mainly due to the fact that we assumed the same $\Delta t$ for all $B_{\rm seed}$ values for simplicity. This means assuming the properties of the turbulence and turbulence driver remains the same for difference seed fields. However, it is possible that these properties are different. For example, in the case of larger $B_{\rm seed}$, the turbulence could be stronger, leading to a quicker amplification of the field (smaller $\Delta t$), and eventually resulting in a smaller $\tau_{\rm dynamo}$.

    These values are upper limits for the e-folding time, as the field could have been amplified quicker than the time between the disk settling into equilibrium and the time when we observe it. The e-folding time could be higher if the disk settled earlier (e.g., $\tau_{\rm dynamo} < 3.8 \times 10^8$~yr, if settled at $z=4.4$, \citealt{2021Sci...372.1201T}). 

    We note that conservation of magnetic helicity provides a lower limit of 1.6 on the ratio of random and coherent fields in mean-field dynamos (\citealt{shukurov_book}, \citealt{2017NatAs...1..621M}). Our results at 6.2\,kpc are consistent with this, as in the case of MW properties we derive a ratio of $\sim$1.6--70. However, the same ratio at 0.7\,kpc is $\sim$0.8--6. This suggests that higher values of turbulent field are more likely to be true (e.g., in the case of a filling factor of $>$0.09, or a turbulent cell size of $<$10 pc), or that the lower end of the derived regular field strengths are more likely (e.g., higher $f_{\rm ion}$).

    \subsection{Comparison to the halo field of nearby galaxies and the Milky Way}
    \label{dis:nearby}
    
    In this section, we compare our derived field strengths at $z~=~0.414$, to those of the Milky Way and nearby (present-day) galaxies.

    The Continuum Halos in Nearby Galaxies -- an EVLA Survey (CHANG-ES, \citealt{2012AJ....144...43I,2019Galax...7...42I}) has conducted observations of nearby edge-on galaxies and derived their magnetic field properties, we list a few of their results in this paragraph. 
    The first observational evidence of regular magnetic fields in the halo of nearby galaxies was measured by \cite{2019A&A...632A..11M}. They found a magnetic field in NGC 4631, which is coherent over 2 kpc in height, and reverses on a scale of 2 kpc along the major axis. The halo field is about 4 $\mu$G in strength, lower than its 9 $\mu$G disk field, which is similar to what we see when we compare our result to another galaxy's disk field at $z=0.4$ (see Section \ref{dis:typical}). 
    \cite{2019Galax...7...54K} found 5--10 $\mu$G total magnetic field strength in the halos of CHANG-ES galaxies. To compare, we 
    estimate the total magnetic field strength from our results, $B_{\rm total} \sim \sqrt{B^2_{\rm random}+B^2_{\rm regular}} = 1 \textrm{--} 7\, \mu$G. This is lower than \cite{2019Galax...7...54K}, but the difference could be partly from the contribution of the anisotropic turbulent field, which we have neglected in the equation above.

    The vertical magnetic field of the Milky Way halo was measured to be $-0.14$ $\mu$G \citep{2009ApJ...702.1230T} or consistent with $0\,\mu$G \citep{2010ApJ...714.1170M} for $h > 0$ (in the northern Galactic hemisphere), and +0.30 $\mu$G for $h < 0$ (in the southern Galactic hemisphere) by exploiting the Faraday rotation of extragalactic radio sources \citep{2009ApJ...702.1230T,2010ApJ...714.1170M}. 
     \cite{2012ApJ...755...21M} assumed that there is a constant, coherent azimuthal halo magnetic field parallel to the Galactic plane, but with different strengths below and above the disk. They calculated a magnetic field strength of 7 $\mu$G below the Galaxy's plane and 2 $\mu$G above it (from 0.8 to 2\,kpc).

    The ranges of these B field strengths are consistent with our result of 0.2--4.1 $\mu$G (see \ref{section:Dis_B}). However, we cannot implement models with more free parameters because we only have one $\Delta$RM available for the lensing galaxy of B1600+434. This is why we assumed that only the direction of the magnetic field can change below and above the disk, not its strength.

    \subsection{Comparison to cosmological MHD simulations}
    
  We selected star-forming galaxies at $z=0.4$ from the TNG50 simulation of the IllustrisTNG project \citep{2019MNRAS.490.3196P,2019MNRAS.490.3234N}, and derived their vertical magnetic field strength profiles from their edge-on magnetic field maps (calculated by integrating through 20--40 kpc depending on the size of a given galaxy). 
    We find that the vertical magnetic field strength is 0.55 $\mu$G at 0.7\,kpc and 0.2 $\mu$G at 6.2\,kpc from the mid-plane. This is consistent with the lower bounds on magnetic field strengths we estimated from our observations. We note that, this comparison between the derived large-scale coherent field strengths from our observations and the magnetic field strengths in the simulation is comparing similar quantities, as TNG50 mainly has a large-scale field and has weaker random magnetic fields compared to observations. This is due to the fact that the simulation does not contain all sources of turbulence on small scales, that could amplify the turbulent field. The reason behind the low large-scale field strengths could be similarly due to the limited spatial resolution of the simulation (i.e., missing sources of turbulence), thus does not completely model the mean-field dynamo seeded by random fields (see discussion in, e.g., \citealt{2024A&A...690A..47K}, Section 3.3.4). 
    
    This is similar to the missing small-scale turbulence leading to lower magnetic field strengths in the Auriga simulations \citep{2017MNRAS.469.3185P,2018MNRAS.481.4410P} compared to observations. Their vertical magnetic field strength profiles show stronger magnetic field strengths than our results from TNG50, with $\sim$2 $\mu$G at $h=6$ kpc and $\sim$3.5 $\mu$G at $h=0.7$ kpc. These are consistent with our results considering the full parameter range, but we derive a $\sim$50\% lower value of $B$ at $h=0.7$ kpc with MW-like parameters. 
    
    \subsection{Typical magnetic fields at $z \simeq 0.4$}
    \label{dis:typical}
    
      \cite{2017NatAs...1..621M} used the same method as we for the lensing system B1152+199, with a lensing spiral galaxy at $z = 0.439$, where the sightlines of the lensed quasar images pass through the disk of the galaxy. They detected a coherent magnetic field with field strengths between 4 and 16 $\mu$G, and found that the scenario of axisymmetric fields in the disk is favored over the presence of bisymmetric fields. Their results are consistent with the dynamo model, because that model predicts that the axisymmetric mode is strongest in the disk (\citealt{2017NatAs...1..621M}, \citealt{2005LNP...664..113S}, \citealt{2009A&A...494...21A}).
    
    The sightlines of B1600+434, on the other hand, pass through the halo of the lensing galaxy, providing information on halo magnetic fields. Assuming that both galaxies are typical spirals at $z \simeq 0.4$, we can draw an overall view of galactic magnetic fields at that redshift, including the halo and the disk. We find that the halo $B$ field strength in the case of the lensing galaxy of B1600+434 being a Milky Way-like galaxy (0.9--2.5 $\mu$G) is lower by a factor of 1.6--19 compared to the disk field (4--17 $\mu$G) in the lensing galaxy of B1152+199 \citep{2017NatAs...1..621M}. In most cases, the halo magnetic field strength is also lower than that in the disk when we consider the full parameter range (0.2--4.1  $\mu$G) for the former. This is also in agreement with dynamo theory.

\section{Conclusions and outlook}
    \label{Conclusion}
    We performed broadband VLA radio polarization observations of the gravitational lensing system B1600+434. We detected RM difference between the lensed images ($\Delta {\rm RM_{\rm rf}}=16 \pm 4 \,\radm$) and differential Faraday dispersion ($\Delta\sigma_{\rm RM,rf}~=~19 \pm 5~\radm$), from which we infer the large-scale and turbulent magnetic field strength in the halo of a spiral galaxy at $z=0.414$.

    If the RM difference between the lensed images is caused by the regular field, assuming Milky-Way-like galaxy properties and an exponential magnetic field model, we find the vertical magnetic field strength to be 1.2--1.8 $\mu$G at a vertical distance of 0.7 kpc from the plane of the galaxy and 0.1--0.9 $\mu$G at 6.2 kpc. With broader parameter ranges and different geometry models, we find values of 0.01--2.8 $\mu$G at 0.7 kpc and 0.2--3 $\mu$G at 6.2 kpc. The strength of the halo field is comparable to halo fields measured in nearby galaxies (\citealt{2015A&ARv..24....4B}, \citealt{2013pss5.book..641B}, CHANG-ES - \citealt{2012AJ....144...43I}) and the Milky Way \citep{2009ApJ...702.1230T,2010ApJ...714.1170M,2012ApJ...755...21M}. We note that in contrast to most magnetic field measurements of nearby galaxies, which are based on the equipartition assumption\footnote{The energy densities of the total cosmic rays and the total magnetic field are equal}, our measurements are independent and not based on that assumption but still show consistent results. This could mean that the mechanism generating the large-scale magnetic field has already built up the regular field strength to what we see in present-day galaxies at a redshift of 0.414. 
    We detected differential Faraday dispersion ($\Delta\sigma_{\rm RM}$) between the lensed images, which is indicative of turbulent magnetic fields on scales lower than 50 pc in the halo of the lensing galaxy. The corresponding field strengths are 1.4--6.9 $\mu$G considering MW-like galaxy properties, and 0.2--12.1 $\mu$G considering a wider range.
    We note that there is a possibility that part of the differential RM is caused by the turbulent field.

    We put our results in context with the dynamo theory. We derive a dynamo e-folding time of < 2.9 $\times$ 10 $^8$ yr, consistent with theory \citep{2009A&A...494...21A}. Dynamo theory predicts that the small-scale turbulent field is stronger than the large-scale field, and that the disk field is stronger than the halo field. We find both of these criteria are fulfilled by our results, as the small-scale turbulent magnetic field  (1.4--6.9 $\mu$G) is stronger in most cases than the large-scale coherent field (1.2--1.8 $\mu$G), and the halo magnetic field in the lensing galaxy is weaker compared to the disk field of a similar galaxy at $z \simeq 0.4$ \citep{2017NatAs...1..621M}.

    We demonstrated that many parameters are important for the accurate derivation of the magnetic field strength in  lensing galaxies, and for excluding the possibility of turbulent fields in the lensing galaxy causing the RM difference. These could be overcome if we have a large enough sample, for example, the SKA is expected to observe $10^5$ new lensing systems (\citealt{2015aska.confE..84M}), which will significantly increase the number of suitable systems for magnetic field studies.

    We note the importance of higher frequency data (e.g., S- and C-band) in detecting the Faraday dispersion caused by the lensing galaxy. As we have seen in Section \ref{Res:RMdiff_other}, observing only in L band can miss polarized emission over broad Faraday depth components (the inability to measure $\sigma_{\rm RM}$). The presence of a polarized component with statistically significant $\sigma_{\rm RM}$ can help us quantify how much turbulent fields contribute to the measured RM difference between sightlines. However, the estimation depends on a few elusive ISM turbulence parameters: 
        the characteristic size of the turbulence that caused the measured $\sigma_{\rm RM}$, the outer scale of turbulence, and how the turbulence scales in the halo of galaxies compared to the disk. Measuring the projected size (e.g., with VLBI) of the background source can provide an estimate on the size of the turbulence causing $\sigma_{\rm RM}$. 
        The observation of multiple similar lensing systems, with similar lensing galaxy redshift, galaxy properties (e.g., spiral, on SF main sequence), and similar inclination, could allow us to construct an RM structure function to characterize the outer scale of the turbulence. Comparing multiple systems would also help to map the large-scale regular magnetic field configuration.

        In a large enough sample, the inclination and electron density of the individual lensing galaxies could become less important. However, as a difference of 20$^{\circ}$ in $i$ can affect the field strength by a factor of $\sim$5 (see Appendix \ref{app:incl_param}), and assumption on $f_{\rm ion}$ can change the results by a factor of 10, obtaining information on these still might be useful. One way to determine $n_e$ is using the differential hydrogen column density (with e.g., CHANDRA), as we did in this work.
        But it can also be achieved by optical and near infrared spectroscopy with the Multi Unit Spectroscopic Explorer (MUSE) and the \textit{James Webb} Space Telescope (JWST), see for example, \citealt{2021A&A...654A..80S,2023ApJ...956..139I}.
        Additionally, we can derive $i$ using optical or near-infrared imaging (HST, JWST, e.g., \citealt{2018ApJ...869..161W,2024MNRAS.530.1984L}).

        In summary, what is needed for a robust future analysis is the projected size of the background sources (VLBI imaging), $i$ (HST or JWST imaging) and $n_e$ of the lensing galaxies (X-ray observations, MUSE or JWST), detection of $\sigma_{\rm RM}$ (broadband radio polarimetry including higher frequency, i.e., S- or C-band), and the joint analysis of many lensing systems. In the future, with lensing galaxies at higher redshifts, we will be able to probe the large-scale dynamo at earlier times. Combining multiple similar systems at the same redshift will help us to characterize the large-scale field geometry and the outer scale of turbulence in distant galaxies.

\begin{acknowledgements}
The National Radio Astronomy Observatory is a facility of the National Science Foundation operated under cooperative agreement by Associated Universities, Inc. We thank the manuscript referee Micha\l{} Hanasz for reviewing the paper and his useful suggestions. We thank Rainer Beck for his useful comments and suggestions. This publication is adapted from part of the PhD thesis of the lead author \citep[Chapter 3]{2024PhDT.........6K}.

\end{acknowledgements}

\bibliographystyle{aa}
\bibliography{aanda}

\appendix

\section{Turbulent magnetic field}
\label{app:turbulent}

We extrapolate $\sigma_{\rm RM}$ from scales smaller than the beam (or projected size of the background source, $\sim$50 pc) to larger scales. We assume that the outer scale of turbulence is 1000 pc (see e.g., turbulence in the halo of M51, \citealt{2015ApJ...800...92M}), thus we estimate the turbulent magnetic field strength and $\sigma_{\rm RM}$ on scales of 1000 pc. To do this, we follow the steps below:

\begin{enumerate}

    \item Derive the turbulent magnetic field strength on small scales (smaller than beam/projected size of background source, which is 50 pc here):
    \begin{equation}
    B_{r,1} = \frac{\sqrt{3fN_1\Delta \sigma_{\rm RM}^2}}{0.812 \sqrt{N_{\rm e,B}^2 -N^2_{\rm e,A}}} = 1.4 - 6.9 \,(0.2 - 12.1)\,\mu {\rm G},
    \end{equation}
     where $\sigma_{\rm RM}$ = 19 $\pm$ 5 $\radm$, the volume filling factor of the warm ionized medium $f = 0.02 $ to $0.14$ (filling factor, \citealt{2008PASA...25..184G}), $N_{\rm 1} = L/l_{\rm 1}$, $L = 10$ kpc and $l_{\rm 1} = 10$ pc. $N_{\rm e,B}$ and $N_{\rm e,B}$ are in the ranges listed in Table \ref{tab:parameter_ranges}. The derived magnetic fields correspond to the MW parameter range, and the full parameter range is in parentheses.

     \item Calculate magnetic field strength on large scales assuming Kolmogorov spectrum:
     \begin{equation}
     B_{r,2} = B_{r,1} \cdot \left(\frac{l_{\rm 1}}{l_{\rm 2}}\right)^\gamma
     \end{equation}
    where $\gamma = -1/3$, and $l_{\rm 2} = 1000$ pc.
    
     \item Add additional scale factor because we are dealing with turbulence in the halo and not the disk, which has a different density. Based on flux freezing \citep{2011hea..book.....L}, we can calculate the magnetic field strength on large-scales in the halo in the following way:
     \begin{equation}
     B_{\rm r,2, halo} = B_{r,2} \cdot \left(\frac{\rho_{\rm halo}}{\rho_{\rm disk}}\right)^n = 0.3 \textrm{--} 1.5 \, (0.05 \textrm{--} 2.6) \, \mu {\rm G},
     \end{equation}
    where $\frac{\rho_{\rm halo}}{\rho_{\rm disk}} = 0.01$ (same assumption as in \citealt{2015A&ARv..24....4B}), and $n = 2/3$. As before, the derived magnetic fields correspond to the MW parameter range, and the full parameter range is in parentheses.

     \item Next, we derive $\sigma_{\rm RM}$ on 1000 pc scale ($\sigma_{\rm RM, 1000}$):
    \begin{multline}
    \sigma_{\rm RM, 1000} = B_{\rm r,2, halo} \frac{0.812 \sqrt{N_{\rm e,B}^2 -N^2_{\rm e,A}}}{\sqrt{3fN_{2}}} = \\ = \sqrt{\Delta \sigma_{\rm RM}^2} \left(\frac{l_{1}}{l_{2}}\right)^{\gamma-1/2} \left(\frac{\rho_{\rm halo}}{\rho_{\rm disk}}\right)^n = 41 \pm 11 \, \radm,
    \end{multline}
    where $N_{\rm 2} = L/l_{\rm 2}$.
    We note that the derived value does not depend on $N_{\rm e,A}$, $N_{\rm e,B}$, $L$, and $f$, as these drop out in the final equation. The parameters that can affect $\sigma_{\rm RM, 1000}$ significantly are the ratio of the halo and disk density, $l_{1}$ and $l_{2}$. For example, if $l_{1}$ would be half, or $l_{\rm 2}$ would be twice as large as the size we assumed, $\sigma_{\rm RM, 1000}$ would almost double. Doubling the ratio of the halo and disk ratio would have an effect of similar magnitude. We also note that the $\sigma_{\rm RM, 1000}$ we derive here is an upper limit, as we assumed that the power-law spectrum extends to arbitrarily large-scales. However, it is possible that in reality, there is a cut-off at a smaller physical scale.
    \item Finally, we can compare $\sigma_{\rm RM, 1000}$ to the rest frame RM difference between images ($\Delta {\rm RM_{\rm rf}}=16 \pm 4 \,\radm$). These have a comparable magnitude, which suggests it is possible that the turbulent field of the lensing galaxy also contributes to the observed RM difference. However, due to its dependence on multiple of our assumptions (turbulence scales and density ratio), $\sigma_{\rm RM, 1000}$ could also be lower. In summary, we note the caveat to our result, that the turbulent field could also contribute to the RM difference, and should be considered in the future (see Section \ref{Conclusion}).

\end{enumerate}

\section{Large-scale field geometry models}
\label{app:geometry}

In this section, we list all the geometry models we use in \ref{section:Dis_B}. We assume that the vertical magnetic field is perpendicular to the plane of the galaxy, and the toroidal field is parallel to the plane of the galaxy. 
    \begin{enumerate}
        \item Constant toroidal magnetic fields in the halo, with the same direction below and above the disk. For this case, the toroidal magnetic field strength is given by:
        \begin{equation}
            B_{\rm t}  = B_{\rm 0} (= B_{\rm A} = B_{\rm B}).
        \end{equation}
        We consider the case where the toroidal magnetic field has no height dependence, and we derive:
        \begin{equation}
            |B_{\rm 0}| = \frac{\rm \Delta RM}{0.812 \sin(i) (N_{\rm e,B}-N_{\rm e,A})} 
        \end{equation}

        \item Constant vertical magnetic fields in the halo, with the same direction below and above the disk (constant, antisymmetric, dipolar). For this case, the vertical magnetic field strength is given by:
        \begin{equation}
            B_{\rm vert}  = B_{\rm 0} (= B_{\rm A} = B_{\rm B}).
        \end{equation}
        In the simplest case, where the magnetic field has the same direction below and above the disk and has no height dependence, we derive:
        \begin{equation}
            |B_{\rm 0}| = \frac{\rm \Delta RM}{0.812 \cos(i) (N_{\rm e,B}-N_{\rm e,A})} 
        \end{equation}

        \item Constant vertical magnetic fields in the halo, with opposite directions below and above the disk (constant, symmetric, quadrupolar). For this case, the vertical magnetic field strength is given by:
        \begin{equation}
            B_{\rm vert}  =
            \begin{cases}
              +B_0 = B_{\rm A}, & \text{if}\ h>0 \\
              -B_0 = B_{\rm B}, & \text{if}\ h<0.
            \end{cases}
        \end{equation}
        
        In this case, 
        \begin{equation}
            |B_{\rm 0}| = \frac{- \rm \Delta RM}{0.812 \cos(i) (N_{\rm e,B}+N_{\rm e,A})}.
        \end{equation}
        
        \item Vertical magnetic field strengths vary with distance from the midplane, with the same direction below and above the disk ($B/h$, antisymmetric). For this case, the vertical magnetic field strength is given by:
        \begin{equation}
           B_{\rm vert} = \frac{B_{\rm 0}}{|h|}.
        \end{equation}
        
        In this case, 
        \begin{equation}
            |B_{\rm 0}| = \frac{\rm \Delta RM}{0.812 \cos(i) \left( \frac{N_{\rm e,B}}{|h_{\rm B}|}+\frac{N_{\rm e,A}}{|h_{\rm A}|}\right)}.
        \end{equation}
        
        \item Vertical magnetic field strengths vary with distance from the midplane, with opposite directions below and above the disk ($B/h$, symmetric). For this case, the vertical magnetic field strength is given by:
        \begin{equation}
            B_{\rm vert} = \frac{B_{\rm 0}}{h}.
        \end{equation}
        
        In this case, 
        \begin{equation}
            |B_{\rm 0}| = \frac{\rm \Delta RM}{0.812 \cos(i) \left( \frac{N_{\rm e,B}}{h_{\rm B}}+\frac{N_{\rm e,A}}{h_{\rm A}}\right)}.
        \end{equation}    
        
        \item Vertical magnetic field strengths varying exponentially with distance from the midplane, with the same direction below and over the disk (exponential, antisymmetric). For this case, the vertical magnetic field strength is given by:
        \begin{equation}
            B_{\rm vert} = B_{\rm 0} \exp\left(- \frac{h}{H_{\rm halo}}\right),
        \end{equation}
        where $H_{\rm halo}$ is the scale height of the halo magnetic field. We note that based on the synchrotron scale height in most nearby edge-on spiral galaxies ($H_{\rm syn}$ = 1--2 kpc), we would expect $H_{\rm halo}$=2--8 kpc \citep{2019Galax...8....4B}. However, we again consider a wider range of parameters (between 2 and 20 kpc) to account for very different environments at large $z$.
        
        In this case, 
        \begin{equation}
            |B_{\rm 0}| = \frac{\rm \Delta RM}{0.812 \cos(i)  \left[N_{\rm e,B} \exp\left(-\frac{h_{\rm B}}{H_{\rm halo}}\right)+ N_{\rm e,A} \exp\left( - \frac{h_{\rm A}}{H_{\rm halo}}\right)\right]}.
        \end{equation}  
        
        \item Vertical magnetic field strengths varying exponentially with distance from the midplane, with opposite signs below and above the disk (exponential, symmetric). For this case, the vertical magnetic field strength is given by:
        \begin{equation}
            B_{\rm vert}  = 
            \begin{cases}
               B_{\rm 0} \exp \left(- \frac{h_{\rm A}}{H_{\rm halo}}\right), & \text{if}\ h>0 \\
              -B_{\rm 0} \exp\left(- \frac{h_{\rm B}}{H_{\rm halo}}\right), & \text{if}\ h<0.
            \end{cases}
        \end{equation}

        In this case, 
        \begin{equation}
            |B_{\rm 0}| = \frac{- \rm \Delta RM}{0.812 \cos(i)  \left[N_{\rm e,B} \exp\left(-\frac{h_{\rm B}}{H_{\rm halo}}\right)- N_{\rm e,A} \exp\left( - \frac{h_{\rm A}}{H_{\rm halo}}\right)\right]}.
        \end{equation}

    \end{enumerate}{}

\section{The effects of the range of assumptions on regular magnetic field strength}
    \label{app:parameters}

        \begin{figure*}
            \centering
            \includegraphics[width=16cm]{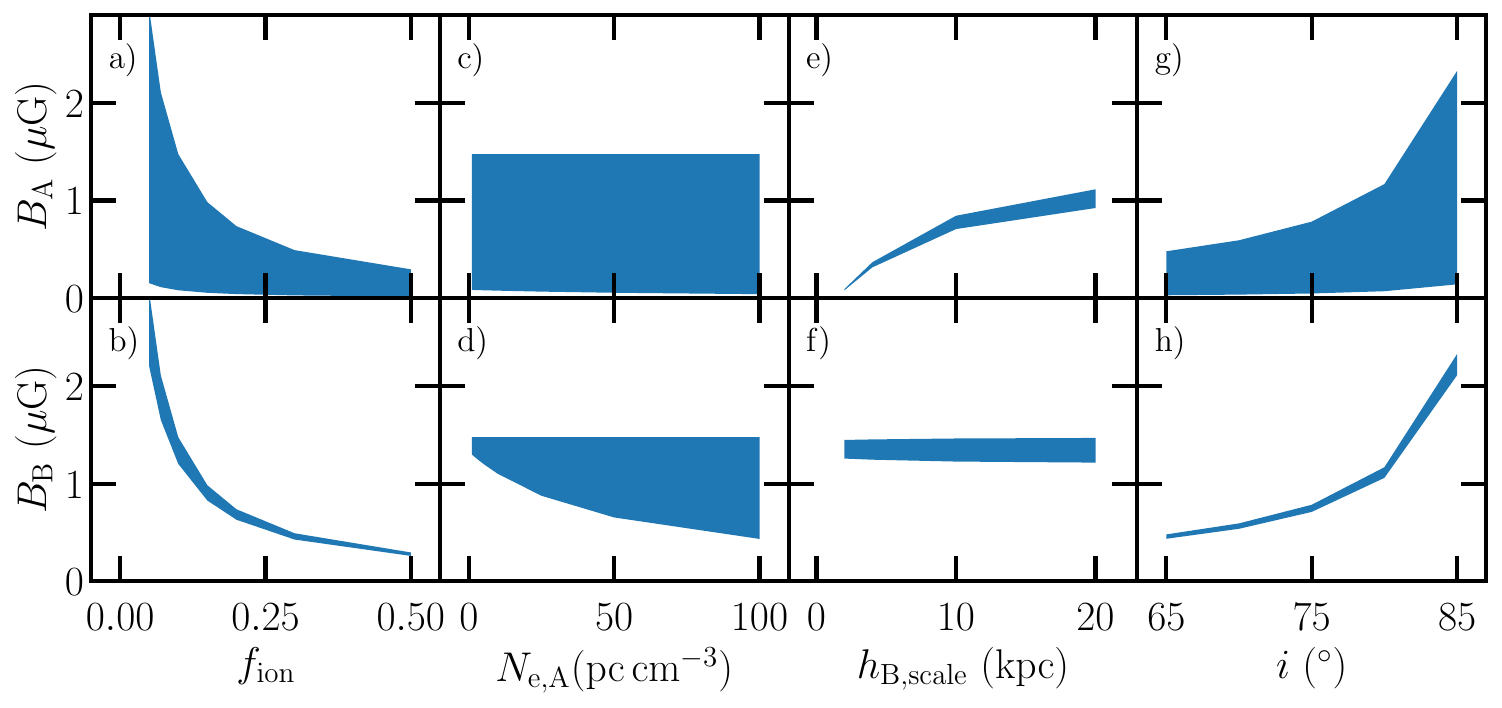}
            \caption{The range of magnetic field strengths in the sightlines of image A and image B calculated with different parameter ranges for ionization fractions (panels a and b), electron column densities assumed at the position of image A (panels c and d), magnetic scale heights (panels e and f), and  inclinations (panels g and h).}
            \label{fig:params}
        \end{figure*}
    
    We investigate how the different assumptions affect our derived B field strength and which parameters affect it the most (see Fig. \ref{fig:params}). We do this by assuming Milky-Way-like galaxy parameters and allowing one parameter to change over its full range. The ranges of these parameters can be found in Table \ref{tab:parameter_ranges}. Below, we list the details of testing different ionization fractions, electron column densities at the position of image A, and scale heights. Apart from these parameters, we also investigated a larger range of inclinations. We used the inclination from \cite{2000ApJ...533..194M} (81.7$^{\circ +0.4^\circ}_{\;-0.5^\circ}$) in our calculation in the main text of the paper and in subsections \ref{app:fion}-\ref{app:zscale}. In subsection \ref{app:incl_param}, we explore a wider range of $65^{\circ}<i<85^{\circ}$. In subsections \ref{app:fion}, \ref{app:NeA}, and \ref{app:incl_param}, we consider all geometry models, and in subsection \ref{app:zscale}, we only consider the exponential models, as only those have exponential scale height in their definition.

        \begin{table}[b]
    \caption{The assumed ranges of the parameters for the lensing galaxy of B1600+434.}

        \centering
        \begin{tabular}{l c c}
        \hline
        \hline
        Parameter & Full range & MW range \\
        \hline
        $f_{\rm ion}$ & 0.05--0.5 & 0.07--0.14 \\
   $\Delta$ $N_e$ ($\dm$) & 48.6--486.1 & 88--98\\
   $N_{\rm e,A}$ ($\dm$) & 1--100 & 2--4 \\
   $N_{\rm e,B}$ ($\dm$) & 49.6--586.1 & 92--100 \\
   $N_{\rm e,A}$ + $N_{\rm e,B}$ ($\dm$) & 50.6--686.1 & 94--104\\
   $H_{\rm halo}$ (kpc) & 2--20 & 2--8\\
   \hline
        \end{tabular}
        \tablefoot{The assumed ranges of the parameters, in a general case (full parameter range) and in the case of Milky Way like galaxy properties (MW parameter range).}
        \label{tab:parameter_ranges}
    \end{table}
        
    \subsection{Ionization fraction}
    \label{app:fion}
     We change the ionization fraction ($f_{\rm ion}$) from 0.05 to 0.5 (shown in panels a and b of Fig. \ref{fig:params}) while keeping the other parameters in the MW parameter range. We find that the magnetic field strength decreases exponentially with increasing ionization fraction (as in the case of higher $f_{\rm ion}$, more of the observed RM difference can be explained by high electron densities instead of strong magnetic fields). If we assume $f_{\rm ion}=0.05$, the derived magnetic field strength can be $\sim$10 times larger than with $f_{\rm ion}=0.5$. We can see this effect even in our MW parameter range; the lowest value of $f_{\rm ion}$ (0.07) can result in a $B$ field strength around double that of the one calculated with its highest value ($f_{\rm ion}=0.014$).

    \subsection{Electron column density at the position of image A}
    \label{app:NeA}
         We vary the electron column density at the position of image A ($N_{\rm e,A}$) over 1--100 $\dm$ (shown in panels c and d of Fig. \ref{fig:params}), while keeping the other parameters in the MW parameter range. We set $f_{\rm ion}$ to 10\%, to exclude changes due to changes in the ionization fraction, since we saw in Appendix \ref{app:fion} that a change in the assumed ionization fraction can significantly affect our derived magnetic field strength. We find that the lower boundary on the range of magnetic field strengths at the position of image B decreases with higher $N_{\rm e,A}$: the $B$ field strength can be three times lower in the case of $N_{\rm e,A}=100 \, \dm$ compared to $N_{\rm e,A}=1 \, \dm$. In the case of the $B$ field strength at the position of image A, we do not find a strong dependence on $N_{\rm e,A}$; here the wide range is due to the different geometry models that give very different $B$ field strengths.
        
    \subsection{Scale height}
    \label{app:zscale}
         We change the scale height ($H_{\rm halo}$) from 2 to 20 kpc (shown in panels e and f of Fig. \ref{fig:params}), while we use the MW parameter range for other parameters, and set $f_{\rm ion}$ to 10\%. At the position of image B (0.7 kpc from the midplane), the scale height does not significantly affect the derived magnetic field strength; it only increases by 1--3\% from $H_{\rm halo} = 2$~kpc to 20 kpc. In the case of the location probed by image A (6.2 kpc away from the midplane), the effect is obviously larger: the magnetic field strength increases by a factor of $\sim$10 from $H_{\rm halo}$ = 2 kpc to 20 kpc.
        
    \subsection{Inclination}
    \label{app:incl_param}
     We explore a wider range of inclinations ($i$) to highlight the magnitude of effect it can have on the magnetic field strength results. 
     While \cite{2000ApJ...533..194M} measured 81.7$^{\circ +0.4^\circ}_{\;-0.5^\circ}$, we obtained $i=70^\circ \pm 3^\circ$ using the \texttt{Smart Inclination Evaluation with Neural Network} \citep{2020ApJ...902..145K}\footnote{\url{https://edd.ifa.hawaii.edu/inclinet/}} on CASTLES images from the NICMOS instrument of the \textit{Hubble} Space Telescope (HST), and \cite{1997ApJ...486..681M} could only constrain it to be larger than 65$^{\circ}$. Thus, we vary $i$ between 65$^{\circ}$ and 85$^{\circ}$ (shown in panels g and h of Fig. \ref{fig:params}), while keeping other parameters in the MW parameter range and setting $f_{\rm ion}$ to 10 \%. We find that the derived magnetic field strengths increase with increasing inclination by a factor of $\sim$5 from $i=65^{\circ}$ to $i=85^{\circ}$.
    
\newpage
\section{The effects of the choice of magnetic field geometry model}
\label{app:Bmodel}
Considering our restricted parameter range of Milky-Way-like parameters, we see that the different models result in different magnetic field strengths (for the summary of values, see Table \ref{tab:B_field_values}). 

The constant fields (both symmetric and antisymmetric) result in (absolute) magnetic field strengths of 1.2--1.8 $\mu$G at the positions probed by both images. The exponential models result in a field strength of 1.2--1.8 $\mu$G at the location of image B (0.7 kpc away from the midplane) and 0.1--0.9 $\mu$G at the location of image A (6.2 kpc away from midplane). The models with simple 1/$h$ scaling also lead to 1.2--1.8 $\mu$G at the position probed by image B, but to a low 0.1--0.2 $\mu$G at the position probed by image A.

We conclude that all the models give consistent results in the case of the location probed by image B. In the case of the location probed by image A, the models give different results, which can be different by a factor of $\sim$10. This difference is similar in magnitude to the variation that different parameter assumptions can cause. All derived magnetic field strengths are of a reasonable magnitude, thus it is not possible to exclude any of the models based on our data.

\end{document}